\documentclass[twocolumn]{aastex631}   

\usepackage{newtxtext,newtxmath}
\usepackage[T1]{fontenc}

\usepackage{nicefrac}
\usepackage{amssymb}
\usepackage{latexsym}
\usepackage{amsmath, mathrsfs, bm}
\usepackage{eqnarray}
\usepackage{url}
\usepackage{cases}
\usepackage{epsfig}
\usepackage{color}
\usepackage{graphicx}
\usepackage{grffile}
\usepackage{float}
\usepackage{soul}
\usepackage{enumerate}
\usepackage{multirow}
\usepackage{threeparttable}
\usepackage{xspace}
\usepackage{bm}

\usepackage{bm}
\usepackage{times}

\usepackage{etoolbox}
\usepackage{multirow}

\begin{document}

\title{Joint Constraints on the Hubble Constant and Cosmic Baryon Density from Localized Fast Radio Bursts and DESI DR2 Baryon Acoustic Oscillations}

\author[0009-0004-5607-9181]{Ji-Guo Zhang}
\affiliation{Liaoning Key Laboratory of Cosmology and Astrophysics, College of Sciences, Northeastern University, Shenyang 110819, China}

\author[0009-0004-6982-4021]{Tian-Nuo Li}
\affiliation{Liaoning Key Laboratory of Cosmology and Astrophysics, College of Sciences, Northeastern University, Shenyang 110819, China}

\author[0009-0005-6921-3201]{Guo-Hong Du}
\affiliation{Liaoning Key Laboratory of Cosmology and Astrophysics, College of Sciences, Northeastern University, Shenyang 110819, China}

\author[0000-0001-5950-7170]{Wan-Peng Sun}
\affiliation{Liaoning Key Laboratory of Cosmology and Astrophysics, College of Sciences, Northeastern University, Shenyang 110819, China}

\author[0000-0002-3512-2804]{Jing-Fei Zhang}
\affiliation{Liaoning Key Laboratory of Cosmology and Astrophysics, College of Sciences, Northeastern University, Shenyang 110819, China}

\author[0000-0002-6029-1933]{Xin Zhang}
\affiliation{Liaoning Key Laboratory of Cosmology and Astrophysics, College of Sciences, Northeastern University, Shenyang 110819, China}
\affiliation{MOE Key Laboratory of Data Analytics and Optimization for Smart Industry, Northeastern University, Shenyang 110819, China}
\affiliation{National Frontiers Science Center for Industrial Intelligence and Systems Optimization, Northeastern University, Shenyang 110819, China}

\correspondingauthor{Tian-Nuo Li}
\email{litiannuo@stumail.neu.edu.cn}
\correspondingauthor{Xin Zhang}
\email{zhangxin@mail.neu.edu.cn}

\begin{abstract}

Fast radio bursts (FRBs) probe free-electron columns along cosmological sightlines through their dispersion measures (DMs), whereas baryon acoustic oscillations (BAO) constrain cosmic expansion geometry through acoustic distance ratios. In this work, we leverage their distinct physical sensitivities to simultaneously determine the Hubble constant ($H_0$) and cosmic baryon density ($\Omega_{\rm b}$) without external cosmic microwave background (CMB) or Big Bang nucleosynthesis (BBN) priors. Combining 120 localized FRBs and DESI DR2 BAO, we obtain $H_0=71.4^{+2.2}_{-2.5}\ {\rm km\ s^{-1}\ Mpc^{-1}}$ and $\Omega_{\rm b}=0.0503^{+0.0025}_{-0.0021}$ in $\Lambda$CDM, corresponding to precisions of $3.3\%$ and $4.6\%$, respectively. The inferred $H_0$ is statistically consistent with both SH0ES and CMB, and $\Omega_{\rm b}$ agrees with the joint constraint from BBN and CMB. Adopting IllustrisTNG-based DM distributions and fixing the diffuse-gas baryon fraction tightens the constraint to $H_0=68.8\pm1.0\ {\rm km\ s^{-1}\ Mpc^{-1}}$, but lowers the central value toward the CMB-inferred value, identifying extragalactic DM modeling as the dominant systematic uncertainty. These results establish FRB-BAO complementarity as a new avenue for simultaneously constraining $H_0$ and $\Omega_{\rm b}$.
\end{abstract}


\section{Introduction}\label{sec:introduction}

A precise late-universe determination of both the Hubble constant ($H_0$) and cosmic baryon density ($\Omega_{\rm b}$) remains challenging. The value of $H_0$ inferred from the cosmic microwave background (CMB) under $\Lambda$CDM differs significantly from local distance-ladder measurements by more than $5\sigma$ \citep{Guo:2018ans,Planck:2018vyg,DiValentino:2019ffd,Vagnozzi:2019ezj,DiValentino:2020zio,DiValentino:2021izs,Vagnozzi:2021gjh,Riess:2021jrx,Abdalla:2022yfr,Schoneberg:2021qvd,Giare:2023xoc,Hu:2023jqc,Poulin:2023lkg,Vagnozzi:2023nrq,Cai:2026swf,Schoneberg:2026vaf,Sabogal:2026ipu}, widely known as the ``Hubble tension''. Meanwhile, many low-redshift baryons are expected to occupy warm, diffuse ionized gas that is difficult to detect through conventional emission or absorption measurements \citep{Fukugita:1997bi,Cen:1998hc,Shull:2011aa}. 
A late-universe probe sensitive to both cosmic expansion and ionized baryons would therefore provide a valuable cross-check of these inferences.

Fast radio bursts (FRBs) provide such a probe. Their dispersion measures (DMs) record the integrated free-electron column density along each sightline, while precise localization supplies the host-galaxy redshift \citep{Lorimer:2007qn,Thornton:2013iua,Chatterjee:2017dqg}. The resulting DM-redshift relation, known as the Macquart relation, has identified the diffuse intergalactic medium (IGM) as a major reservoir of missing baryons \citep{McQuinn:2013tmc,Macquart:2020lln,Yang:2022ftm,Connor:2024mjg}. Early studies established the cosmological potential of FRBs \citep{Deng:2013aga,Zhou:2014yta,Gao:2014iva,Yang:2016zbm}. The Macquart relation has since been widely used to infer the Hubble constant \citep{Hagstotz:2021jzu,Wu:2021jyk,James:2022dcx,Zhao:2022yiv,Wei:2023avr,Kalita:2024xae,Piratova-Moreno:2025cpc,Acharya:2025ubt,Gao:2025fcr,Xu:2025ddk,Zhuge:2025urk,Liu:2026epv} and dark energy \citep{Walters:2017afr,Zhao:2020ole,Qiu:2021cww,Wu:2022dgy,Zhao:2022bpd,Zhang:2023gye,Dai:2023ska}. More recently, FRB DM statistics and cross-correlations with large-scale structure have opened new avenues for probing cosmic baryon distributions and baryonic feedback \citep{Wang:2025ibv,Wang:2026llr,Sharma:2026xbj,Wayland:2026dhj,Leung:2025vej}. For recent reviews of FRB physics and cosmological applications, see \citet{Zhang:2022uzl,Wu:2024iyu,Wang:2026tgi}.

Cosmological inference from the Macquart relation is, however, intrinsically degenerate \citep{Kumar:2019qhc}. The mean intergalactic contribution scales as $\langle{\rm DM}_{\rm IGM}\rangle\propto\Omega_{\rm b}H_0f_{\rm d}$, where $f_{\rm d}$ is the fraction of baryons in diffuse ionized gas. Consequently, many direct FRB determinations of $H_0$ have fixed $\Omega_{\rm b}$ or adopted priors informed by the CMB or Big Bang nucleosynthesis (BBN) \citep{Hagstotz:2021jzu,Wu:2021jyk,James:2022dcx,Zhao:2022yiv,Wei:2023avr,Wang:2025ugc,Piratova-Moreno:2025cpc,Zhuge:2025urk}. Conversely, FRB-based determinations of $\Omega_{\rm b}$ have generally relied on an assumed $H_0$ or an external $H_0$ prior \citep{Macquart:2020lln,Yang:2022ftm}. Thus, although FRBs are sensitive to both parameters, the product dependence prevents a precise simultaneous determination of $H_0$ and $\Omega_{\rm b}$ from FRBs alone. To overcome this limitation, previous analyses have supplemented FRBs with direct measurements of $H(z)$, supernova distances, gravitational-wave standard sirens, and strong-lensing time delays \citep{Liu:2022bmn,Gao:2023izj,Fortunato:2024hfm,Zhang:2024rra,Zhang:2025yhi,Jahns-Schindler:2025saf,Sales:2026smc,Wu:2026mox}. 

Recently, the Dark Energy Spectroscopic Instrument (DESI) Data Release 2 (DR2) has delivered high-precision BAO measurements over a broad redshift range \citep{DESI:2025zgx,DESI:2025zpo}. 
When combined with CMB observations and type Ia supernovae (SNe) data, DESI DR2 places more stringent constraints on the dark energy equation of state parameters, revealing a $2.8\sigma$--$4.2\sigma$ preference for dynamical dark energy and opening new avenues for cosmological investigations \citep{Jiang:2024xnu,Giare:2024smz,Giare:2024gpk,Lemos:2025bgy,Pang:2025lvh,Ye:2024ywg,Liu:2025mub,Li:2025ops,You:2025uon,Li:2026xaz,Li:2026hwq,Du:2026qtq,Li:2025vuh,Li:2025owk,Turner:2026vzp,Ribeiro:2026lyd,Bella:2026zuk,Wang:2024dka,Giare:2026oti,Wang:2026wrk,Wang:2025znm,Yang:2025gaz,Cai:2025mas,Ren:2026jyw,Wang:2026vqw,Yang:2026qmz,Pan:2025qwy,Yang:2026yaq,Du:2025xes,Pedrotti:2026ncd,Yao:2025twv,Mukherjee:2026lnb,Yang:2026vti,Fortunato:2025qxc}; see also a recent review \citep{Li:2026asg}. 
In particular, \citet{Wang:2025ugc} combined 92 localized FRBs with DESI DR1 BAO and PantheonPlus SNe data in a joint analysis of dynamical dark energy, finding a $2.5\sigma$ preference for a time-varying equation of state even without CMB data.

In this work, we combine 120 localized FRBs with DESI DR2 BAO to obtain a simultaneous late-universe determination of $H_0$ and $\Omega_{\rm b}$. We find that the FRB and BAO constraints exhibit nearly orthogonal degeneracy directions in the $H_0$--$\Omega_{\rm b}$ plane, allowing their combination to efficiently break the individual parameter degeneracies. The joint analysis yields $H_0=71.4^{+2.2}_{-2.5}\ {\rm km\,s^{-1}\,Mpc^{-1}}$ and $\Omega_{\rm b}=0.0503^{+0.0025}_{-0.0021}$, with precisions of $3.3\%$ and $4.6\%$, respectively, demonstrating the strong potential of the FRB-BAO combination.

\section{Data and Methods}

\subsection{Localized FRB Sample}\label{sec:localized_frb_sample}

We compile an initial sample of 139 localized FRBs.
For a localized FRB at redshift $z_i$, the observed dispersion measure is decomposed as
\begin{equation} {\rm DM}_{{\rm obs},i} = {\rm DM}_{{\rm ISM},i} + {\rm DM}_{{\rm halo},i} + {\rm DM}_{{\rm IGM},i} + \frac{{\rm DM}_{{\rm host},i}}{1+z_i}, \label{eq:dmdec} \end{equation}
where the four terms denote the contributions from the Milky Way interstellar-medium (ISM), Milky Way halo, IGM, and host environment, respectively, with the host term defined in the source frame.
In the baseline analysis, we estimate and subtract the Milky Way ISM contribution using the NE2001 model \citep{Cordes:2002wz}, defining the excess DM as
\begin{equation}
{\rm DM}_{{\rm E},i}\equiv {\rm DM}_{{\rm obs},i}-{\rm DM}_{{\rm ISM},i}.
\end{equation}
We retain sources with finite values of $z$, ${\rm DM}_{\rm obs}$, and ${\rm DM}_{\rm ISM}$ that satisfy
\begin{equation} \rm DM_E>80\,\mathrm{pc\,cm^{-3}},\qquad \rm DM_{\rm ISM}\leq0.4\,\rm DM_{\rm obs}. \label{eq:quality_cuts} \end{equation}
These cuts remove sightlines for which the Galactic subtraction dominates the measured column or leaves little extragalactic DM \citep{Connor:2024mjg}. We additionally exclude FRB~20190520B and FRB~20220831A due to excess local DM contributions from their host environments. The final sample contains 120 FRBs: 74 at $z<0.3$, 40 at $0.3\leq z<0.7$, and 6 at $z\geq0.7$. It spans $0.0065\leq z\leq2.148$, has a median redshift of $0.2418$, and includes FRB~20240304B as its highest-redshift event. Appendix~\ref{sec:frb_sample} lists the complete 120-event analysis catalog.  Recent forward-modeling studies suggest that selection effects introduce relatively small cosmological parameter biases ($\lesssim0.8\sigma$) in conditional DM--redshift analyses of current samples of $\sim100$ FRBs, but may become significant for future samples of $\sim10^4$ FRBs \citep{Sharma:2025lql}.

\subsection{FRB DM Model}
The Milky Way halo contribution is marginalized over a Gaussian prior with $\mu_{\rm halo}=65\ {\rm pc\ cm^{-3}}$ and $\sigma_{\rm halo}=15\ {\rm pc\ cm^{-3}}$. We adopt this Gaussian form as a modeling choice, motivated by Galactic-halo estimates of approximately $50$--$80\ {\rm pc\ cm^{-3}}$ \citep{Prochaska_2019}. To assess Milky Way foreground systematics, we replace NE2001 with YMW16 \citep{Yao:2017kcp} and NE2025 \citep{Ocker:2026mta} for the ISM contribution, and the Gaussian halo model with the sightline-dependent YT2020 \citep{Yamasaki:2019htx} and Huang25 \citep{Huang:2024hgz} predictions. YMW16, NE2025, and YT2020 are evaluated using \textsc{PyGEDM} \citep{Price:2021gzo},\footnote{\url{https://github.com/FRBs/pygedm}} while Huang25 is implemented with \textsc{pyhesdm}.\footnote{\url{https://github.com/FRBs/pyhesdm}} Recent FRB-based studies have also investigated possible anisotropy in the Galactic-halo DM contribution \citep{Liu:2026txt,Wang:2026halo}.

The cosmological signal enters through the mean IGM DM \citep{Macquart:2020lln,Ioka:2003fr,Inoue:2003ga},
\begin{equation} \left<{\rm DM}_{\rm IGM}(z)\right> =\frac{3c\Omega_{\rm b}H_0f_{\rm d}} {8\pi Gm_{\rm p}} \int_0^z \frac{\chi_{\rm e}(z')(1+z')}{E(z')}dz'. \label{eq:meanigm} \end{equation}
Here, $E(z)=H(z)/H_0$ is the dimensionless Hubble parameter, $c$ is the speed of light, and $m_{\rm p}$ is the proton mass. The number of free electrons per baryon is $\chi_{\rm e}=Y_{\rm H}X_{\rm e,H}+Y_{\rm He}X_{\rm e,He}/2$, where $X_{\rm e,H}$ and $X_{\rm e,He}$ are the ionization fractions of hydrogen and helium, respectively. Assuming fully ionized hydrogen and helium at $z<3$, with $Y_{\rm H}=0.75$ and $Y_{\rm He}=0.25$, gives $\chi_{\rm e}=7/8$. Equation~(\ref{eq:meanigm}) shows that the mean IGM dispersion measure scales as $\Omega_{\rm b}H_0f_{\rm d}$. Given the uncertainties in $f_{\rm d}$, previous studies have modeled the diffuse baryon fraction as either constant or redshift-dependent \citep{Li:2019klc,Wei:2019uhh,Li:2020qei,Wang:2022ami,Liu:2025fdf}. Recent FRB observations have also provided constraints on the cosmic baryon distribution and its partition between the IGM and galactic halos \citep{Connor:2024mjg,Khrykin:2024lmd}.
To account for the degeneracy between $\Omega_{\rm b}H_0$ and $f_{\rm d}$, we marginalize over $f_{\rm d}$ in our baseline analysis using a flat prior of $f_{\rm d}\in[0.84,0.96]$, following \citet{Connor:2024mjg}. This prior is derived from FRB-independent observations of the baryon fractions in stars ($f_*$) and cold gas ($f_{\rm cold}$), through $f_{\rm d}=1-f_*-f_{\rm cold}$. To conservatively account for uncertainties in the stellar baryon census, the adopted prior allows $f_*$ to span a factor of five.

Line-of-sight fluctuations in the cosmic web are modeled with the dimensionless variable
\begin{equation} \Delta\equiv \frac{{\rm DM}_{\rm IGM}}{\left<{\rm DM}_{\rm IGM}\right>}. \end{equation}
Following the standard Macquart-relation likelihood \citep{Macquart:2020lln}, we adopt
\begin{equation} P_{\rm IGM}(\Delta)= A\Delta^{-\beta} \exp\left[ -\frac{(\Delta^{-\alpha}-C_0)^2} {2\alpha^2\sigma_{\rm IGM}^2} \right], \quad \Delta>0, \label{eq:pigm} \end{equation}
with $\alpha=\beta=3$ and $\sigma_{\rm IGM}=Fz^{-1/2}$, where $F$ is the parameter reflecting the galactic feedback. The constants $A$ and $C_0$ enforce normalization and $\langle\Delta\rangle=1$.

The host contribution is described by a lognormal distribution in the host frame,
\begin{equation} P_{\rm host}({\rm DM}_{\rm host}) =\frac{1}{\sqrt{2\pi}\sigma_{\rm host}{\rm DM}_{\rm host}} \exp\left[ -\frac{(\ln{\rm DM}_{\rm host}-\mu_{\rm host})^2} {2\sigma_{\rm host}^2} \right]. \label{eq:phost} \end{equation}
As in Equation~(\ref{eq:dmdec}), the observer-frame contribution is ${\rm DM}_{\rm host}/(1+z)$. The mean value and variance of the distribution are ${\rm e}^{\mu_{\rm host}}$ and ${\rm e}^{2\mu_{\rm host}+\sigma_{\rm{host}}^2}({\rm e}^{\sigma_{\rm{host}}^2}-1)$, respectively. For each burst, the likelihood is the convolution
\begin{align} P_i({\rm DM}_{{\rm E},i}\mid\bm{\theta}) ={}& \int {\rm d}{\rm DM}_{\rm halo}\, P_{\rm halo}({\rm DM}_{\rm halo}) \nonumber\\ &\times \int {\rm d}{\rm DM}_{\rm host}\, P_{\rm host}({\rm DM}_{\rm host}) \frac{P_{\rm IGM}(\Delta_i)} {\left\langle{\rm DM}_{\rm IGM}(z_i)\right\rangle}, \label{eq:frbconv} \end{align}
where
\begin{equation} \Delta_i= \frac{{\rm DM}_{{\rm E},i} -{\rm DM}_{\rm halo} -{\rm DM}_{\rm host}/(1+z_i)} {\left<{\rm DM}_{\rm IGM}(z_i)\right>}. \end{equation}
Only the domain $\Delta_i>0$ contributes to the integral. The total FRB likelihood is
\begin{equation} \ln\mathcal{L}_{\rm FRB} =\sum_{i=1}^{N_{\rm FRB}}\ln P_i, \quad N_{\rm FRB}=120. \end{equation}
We also consider redshift-dependent $P_{\rm IGM}$ and $P_{\rm host}$ distributions calibrated with IllustrisTNG (TNG) \citep{Pillepich:2017jle,Nelson:2018uso,Zhang:2020mgq,Zhang:2020xoc}.

\begin{table}[htbp]
\centering
\caption{
Uniform prior ranges for the cosmological and FRB parameters sampled in this analysis. The neutrino mass is fixed at $\sum m_\nu = 0.06\,{\rm eV}$.
}
\label{table1}
\renewcommand{\arraystretch}{1.35}
\begin{tabular}{ccc}
\hline\hline
\textbf{Model/Data} & \textbf{Parameter} & \textbf{Prior} \\
\hline
& $\Omega_{\rm m}$ & $\mathcal{U}[0.1,0.9]$ \\
& $\Omega_{\rm b}$ & $\mathcal{U}[0.01,0.10]$ \\
& $H_0\,[{\rm km\,s^{-1}\,Mpc^{-1}}]$ & $\mathcal{U}[20,140]$ \\
$\Lambda$CDM model & $n_\mathrm{s}$ & $\mathcal{U}[0.8,1.2]$ \\
& $\log(10^{10}A_\mathrm{s})$ & $\mathcal{U}[1.61,3.91]$ \\
& $\tau_\mathrm{reio}$ & $\mathcal{U}[0.01,0.8]$ \\
\hline
& $f_\mathrm{d}$ & $\mathcal{U}[0.84,0.96]$ \\
FRB data & $F$ & $\mathcal{U}[0.011,0.5]$ \\
& $e^{\mu_\mathrm{host}}~[\rm pc ~cm^{-3}]$ & $\mathcal{U}[20,200]$ \\
& $\sigma_\mathrm{host}$ & $\mathcal{U}[0.2,2.0]$ \\
\hline\hline
\end{tabular}
\end{table}

\begin{table*}[htbp] 
\caption{
Cosmological constraints from 120 localized FRBs and DESI DR2 BAO data in the $\Lambda$CDM and $w_0w_a$CDM models. For $\Lambda$CDM, we report the individual and joint constraints, together with tests using external priors. The $w_0w_a$CDM analysis additionally includes SNe data from PantheonPlus, DES-Dovekie, and DESY5. Quoted uncertainties denote $68\%$ credible intervals, while one-sided bounds are given at $95\%$ credibility.
}\label{table2}
\vspace{-5mm}
\setlength{\tabcolsep}{1.0 mm} \renewcommand{\arraystretch}{1.5} 
\begin{center}
\small
\begin{tabular}{lcccccc}
\hline\hline
Model/Dataset & $\Omega_{\rm m}$ & $\Omega_{\rm b}$ & $H_0\ [\mathrm{km\,s^{-1}\,Mpc^{-1}}]$ & $w_0$ & $w_a$ & $r_{\rm d}\ [\mathrm{Mpc}]$ \\
\hline
\multicolumn{7}{l}{\textbf{$\bm{\Lambda}$CDM}} \\
FRB & $<0.2154$ & $0.0531\pm0.0224$ & $75.0^{+20.0}_{-40.0}$ & --- & --- & $163.0^{+50.0}_{-40.0}$ \\
DESI DR2 & $0.2971\pm0.0088$ & $0.0504^{+0.0135}_{-0.0030}$ & $90.0^{+30.0}_{-40.0}$ & --- & --- & $124.0^{+40.0}_{-50.0}$ \\
FRB + DESI DR2 & $0.2967\pm0.0085$ & $0.0503^{+0.0025}_{-0.0021}$ & $71.4^{+2.2}_{-2.5}$ & --- & --- & $142.4\pm4.8$ \\
FRB + $\Omega_{\rm b}h^2$ (BBN prior) & $<0.2343$ & $0.0522^{+0.0082}_{-0.0108}$ & $66.0\pm6.3$ & --- & --- & $165.8^{+15.7}_{-11.3}$ \\
FRB + $H_0$ (CMB prior) & $<0.2207$ & $0.0496^{+0.0037}_{-0.0050}$ & $67.1\pm0.5$ & --- & --- & $164.7^{+16.5}_{-12.3}$ \\
FRB + $H_0$ (SH0ES prior) & $<0.2011$ & $0.0459\pm0.0041$ & $72.9^{+1.1}_{-1.2}$ & --- & --- & $160.3^{+16.6}_{-10.4}$ \\
\hline
\multicolumn{7}{l}{\textbf{$\bm{w_0w_a}$CDM}} \\
FRB + DESI DR2 & $0.3445^{+0.0428}_{-0.0240}$ & $0.0557\pm0.0043$ & $68.5^{+2.9}_{-3.7}$ & $-0.551^{+0.363}_{-0.214}$ & $<-1.03$ & $140.0^{+5.4}_{-6.9}$ \\
FRB + DESI DR2 + PantheonPlus & $0.3014^{+0.0210}_{-0.0114}$ & $0.0530^{+0.0030}_{-0.0037}$ & $69.2\pm3.7$ & $-0.887^{+0.057}_{-0.065}$ & $-0.22\pm0.44$ & $144.5^{+6.3}_{-8.0}$ \\
FRB + DESI DR2 + DES-Dovekie & $0.3115^{+0.0181}_{-0.0119}$ & $0.0519^{+0.0027}_{-0.0034}$ & $70.4\pm3.4$ & $-0.846^{+0.067}_{-0.076}$ & $-0.54\pm0.48$ & $141.8^{+5.9}_{-7.5}$ \\
FRB + DESI DR2 + DESY5 & $0.3200^{+0.0158}_{-0.0113}$ & $0.0527\pm0.0029$ & $70.2\pm3.2$ & $-0.780^{+0.066}_{-0.074}$ & $-0.76\pm0.45$ & $140.8\pm6.3$ \\
\hline\hline
\end{tabular}
\end{center}\end{table*}

\subsection{BAO, SNe, and Cosmological Likelihood}

For a spatially flat universe, the transverse comoving distance and Hubble distance are
\begin{equation} D_M(z)=\frac{c}{H_0}\int_0^z\frac{dz'}{E(z')}, \quad D_H(z)=\frac{c}{H(z)}. \end{equation}
The volume-averaged BAO distance is
\begin{equation} D_V(z)=\left[zD_M^2(z)D_H(z)\right]^{1/3}. \end{equation}
DESI DR2 provides measurements of $D_M/r_{\rm d}$, $D_H/r_{\rm d}$, and/or $D_V/r_{\rm d}$ across multiple redshift bins. The drag-epoch sound horizon is
\begin{equation} r_{\rm d}=\int_{z_{\rm d}}^\infty\frac{c_s(z)}{H(z)}\,dz, \label{eq:rd} \end{equation}
where $c_s(z)$ is the photon-baryon sound speed. We impose no external CMB or BBN priors on $H_0$ or $\Omega_{\rm b}h^2$. Instead, the assumed background model determines $r_{\rm d}$ self-consistently from the sampled physical densities. 

The BAO likelihood is written as
\begin{equation} -2\ln\mathcal{L}_{\rm BAO} = \left[\bm{d}_{\rm BAO}-\bm{m}_{\rm BAO}(\bm{\theta})\right]^T \bm{C}_{\rm BAO}^{-1} \left[\bm{d}_{\rm BAO}-\bm{m}_{\rm BAO}(\bm{\theta})\right], \end{equation}
where $\bm{C}_{\rm BAO}$ is the DESI DR2 covariance matrix. For the dynamical dark-energy analysis, we include SNe likelihoods from PantheonPlus, DESY5, and DES-Dovekie \citep{Brout:2022vxf,Scolnic:2021amr,DES:2024jxu,DES:2025sig}. The total likelihood is
\begin{equation} \ln\mathcal{L}_{\rm tot} =\ln\mathcal{L}_{\rm FRB} +\ln\mathcal{L}_{\rm BAO} +\ln\mathcal{L}_{\rm SNe}, \end{equation}
with the final term omitted for the baseline FRB+DESI constraints.

\begin{figure*}[t]
\centering
\includegraphics[width=0.93\textwidth]{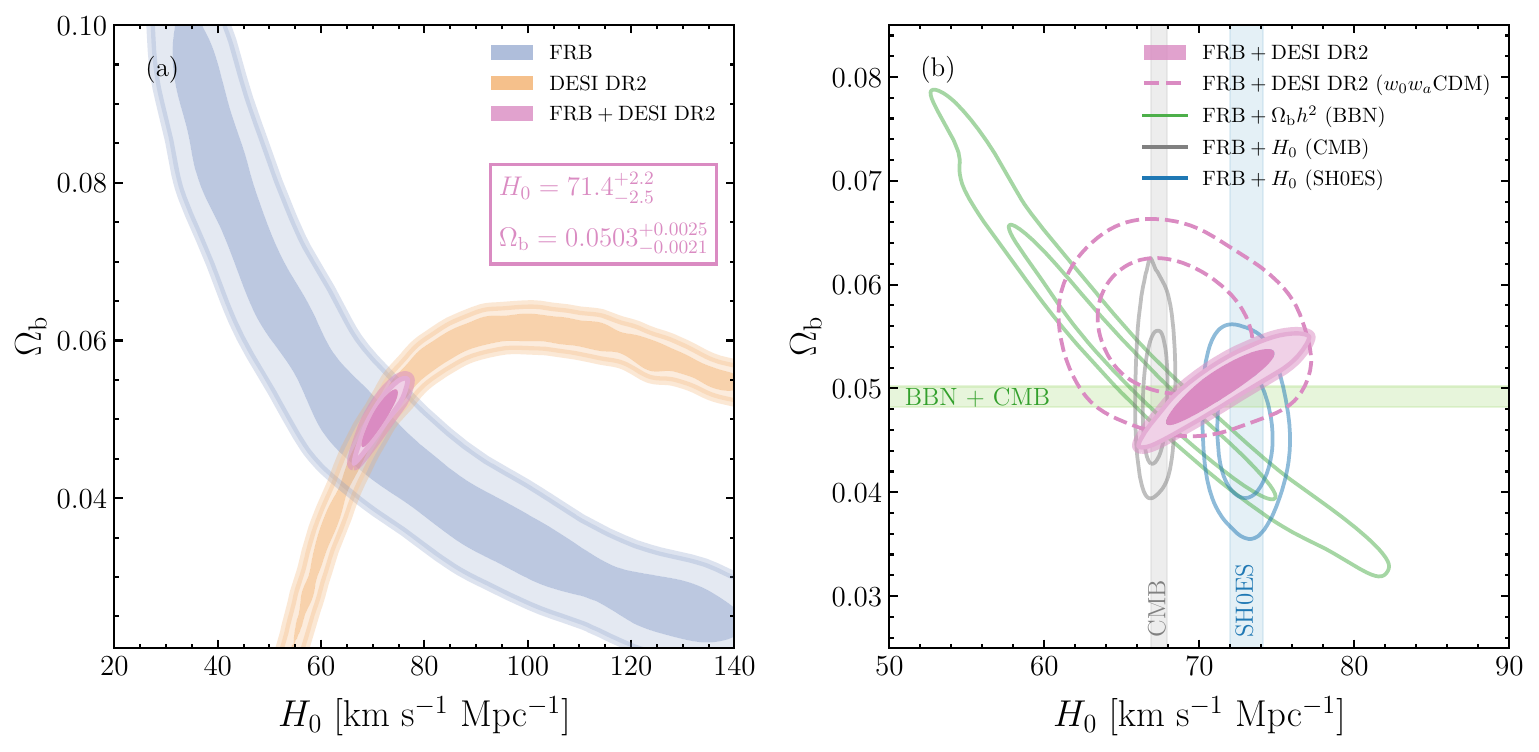}
\caption{
Constraints in the $H_0$--$\Omega_{\rm b}$ plane. Left: FRB, DESI DR2, and joint FRB+DESI DR2 contours in $\Lambda$CDM. Right: comparison with external BBN \citep{Schoneberg:2024ifp}, CMB \citep{Planck:2018vyg}, and SH0ES \citep{Riess:2021jrx} priors, together with the $w_0w_a$CDM extension.}
\label{fig:h0ob}
\end{figure*}

In $\Lambda$CDM,
\begin{equation} E^2(z)=\Omega_{\rm m}(1+z)^3+1-\Omega_{\rm m}, \end{equation}
where $\Omega_{\rm m}$ is the matter density parameter. For the $w_0w_a$CDM parameterization \citep{Chevallier:2000qy,Linder:2002et},
\begin{equation} 
w(a)=w_0+w_a(1-a), 
\end{equation}
and
\begin{align} E^2(z)=& ~\Omega_{\rm m}(1+z)^3 + \left(1-\Omega_{\rm m}\right) \nonumber\\ &\times (1+z)^{3(1+w_0+w_a)} \exp\left[-\frac{3w_a z}{1+z}\right]. \end{align}
The baseline $\Lambda$CDM analysis samples ten cosmological, baryonic, and FRB nuisance parameters listed in Table~\ref{table1}. The $w_0$$w_a$CDM model analysis additionally includes $w_0$ and $w_a$. We adopt uniform priors on all parameters over the ranges. Here, $F$ controls the width of the IGM DM distribution, while $\mu_{\rm host}$ and $\sigma_{\rm host}$ specify the location and width of the lognormal host-DM distribution. The sound horizon $r_{\rm d}$ is derived from the sampled cosmological parameters within the assumed background model and is not assigned an independent prior. We perform Markov chain Monte Carlo analyses using \texttt{Cobaya}~\citep{Torrado:2020dgo}, requiring the Gelman--Rubin convergence criterion $R-1<0.02$~\citep{Gelman:1992zz}. The resulting chains are analyzed using \texttt{GetDist}~\citep{Lewis:2019xzd}.

\section{Results and Discussion}\label{sec:results}

\subsection{The $H_0$--$\Omega_{\rm b}$ Orthogonality}

We summarize the cosmological parameter constraints in Table~\ref{table2}. Using FRB data alone yields a relatively broad constraint on $H_0$ and $\Omega_{\rm b}$, approximately following a constant $\Omega_{\rm b}H_0f_{\rm d}$ in Figure~\ref{fig:h0ob}(a). As follows from Equation~(\ref{eq:meanigm}), at fixed $E(z)$ and $f_{\rm d}$, the leading response of the mean IGM dispersion measure is
\begin{equation} \delta\ln\langle{\rm DM}_{\rm IGM}\rangle \simeq \delta\ln H_0+\delta\ln\Omega_{\rm b}. \end{equation}
A larger $H_0$ can therefore be compensated by a smaller $\Omega_{\rm b}$. This trade-off produces the strong negative degeneracy direction. 

Incorporating DESI DR2 significantly breaks this degeneracy.
Without an external calibration of $r_{\rm d}$ (e.g., provided by the CMB), BAO measurements act as an uncalibrated ruler and constrain the combination $H_0r_{\rm d}$ rather than $H_0$ and $r_{\rm d}$ individually. In the $H_0$--$\Omega_{\rm b}$ plane, this degeneracy translates into a broad constraint band that is nearly transverse to the FRB degeneracy. The near-orthogonality reflects the different physical information encoded in the two probes: FRBs probe the column density of ionized matter through the dispersion measure, whereas BAO constrain cosmic geometry relative to the acoustic ruler. Their combination therefore yields a compact joint posterior,
\begin{equation} \left. \begin{aligned} H_0 &= 71.4^{+2.2}_{-2.5}\ {\rm km\ s^{-1}\ Mpc^{-1}},\\ \Omega_{\rm b} &= 0.0503^{+0.0025}_{-0.0021}, \end{aligned} \right\} \begin{aligned} &{\rm FRB}\\ &+{\rm DESI\ DR2\ BAO}, \end{aligned} \end{equation}
with a corresponding sound-horizon constraint of $r_{\rm d}=142.4\pm4.8\ {\rm Mpc}$. These constraints correspond to $3.3\%$ precision in both $H_0$ and $r_{\rm d}$ and $4.6\%$ precision in $\Omega_{\rm b}$.
In Figure~\ref{fig:h0ob}(b), the inferred value of $H_0$ lies between the CMB value of $H_0=67.36\pm0.54\ {\rm km\,s^{-1}\,Mpc^{-1}}$ and the local distance-ladder value of $H_0=73.04\pm1.04\ {\rm km\,s^{-1}\,Mpc^{-1}},$ and is statistically consistent with both, at $1.7\sigma$ and $0.6\sigma$, respectively.
The constraint is statistically tighter than the TDCOSMO-2025 determination, $H_0=71.6^{+3.9}_{-3.3}\,\mathrm{km\,s^{-1}\,Mpc^{-1}}$, obtained from eight time-delay lenses combined with PantheonPlus constraints on $\Omega_{\rm m}$ \citep{TDCOSMO:2025dmr}.
For $\Omega_{\rm b}$, it exhibits good concordance with the joint BBN + CMB constraint at better than $1\sigma$.
We also examine the inclusion of SNe and find negligible improvement in the constraints on $H_0$ and $\Omega_{\rm b}$, due to their limited ability to break the remaining parameter degeneracies.

We compare our baseline FRB + DESI DR2 BAO results with determinations relying on external priors in Figure~\ref{fig:h0ob}(b). Since FRBs constrain the combination $\Omega_{\rm b}H_0f_{\rm d}$, individual constraints on $\Omega_{\rm b}$ and $H_0$ are highly sensitive to the adopted priors. Applying a recent BBN prior of $0.02218 \pm 0.00055$ on $\Omega_{\rm b}h^2$ \citep{Schoneberg:2024ifp} gives $H_0=66.0 \pm 6.3\ {\rm km\ s^{-1}\ Mpc^{-1}}$, whereas the CMB prior on $H_0$ yields $\Omega_{\rm b}=0.0496^{+0.0037}_{-0.0050}$, and the SH0ES prior on $H_0$ yields $\Omega_{\rm b}=0.0459 \pm 0.0041$. In comparison, the FRB + DESI DR2 BAO combination provides tighter simultaneous constraints on both $H_0$ and $\Omega_{\rm b}$ without relying on external priors for either parameter.

\subsection{Dynamical Dark Energy}

\begin{figure}[htb]
\centering
\includegraphics[width=0.98\linewidth]{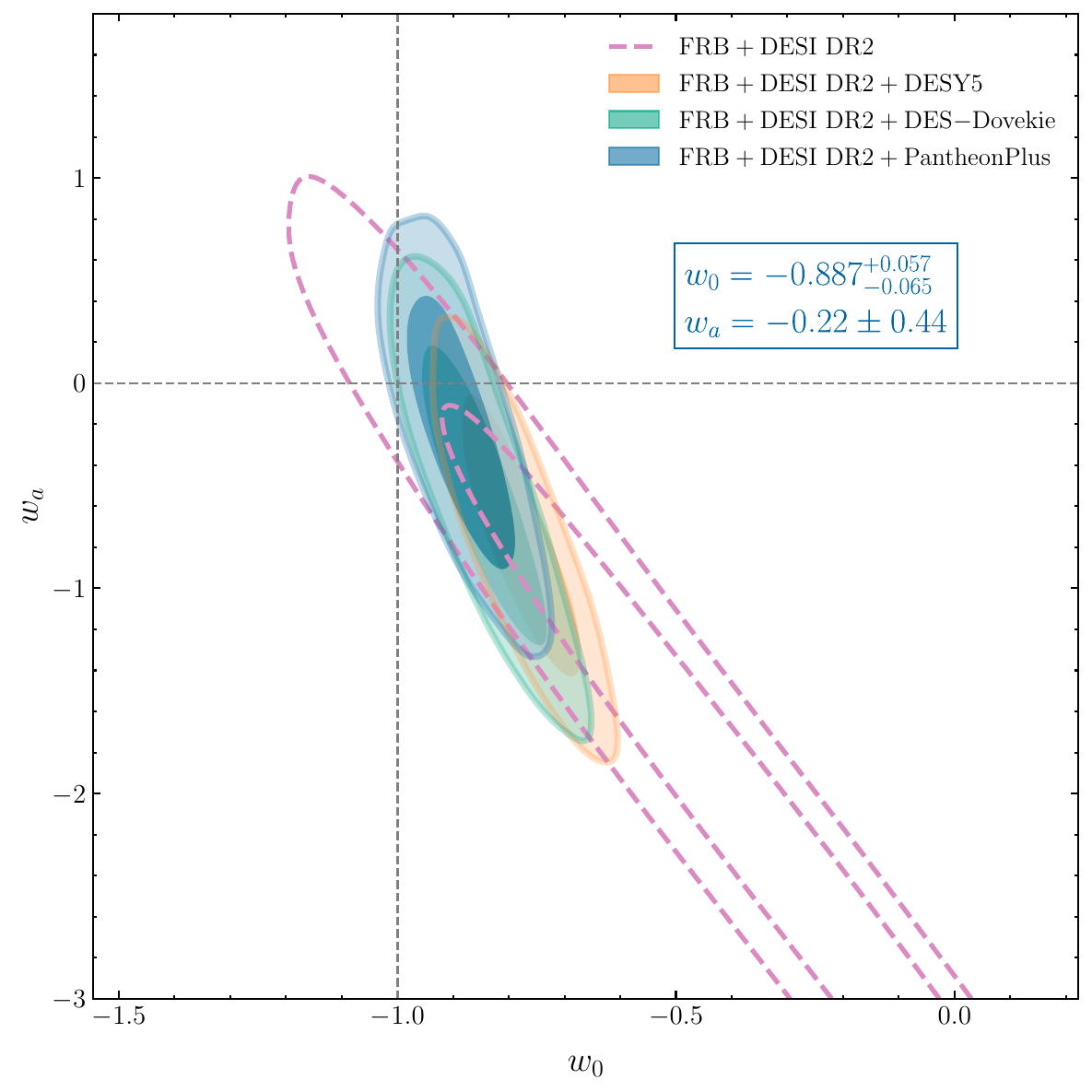}
\caption{
Constraints on the $w_0w_a$CDM dark-energy parameters. FRB + DESI DR2 BAO alone gives a broad degeneracy, while the addition of PantheonPlus, DES-Dovekie, or DESY5 SNe distances significantly tightens the $w_0$--$w_a$ constraints. The dashed gray lines mark the $\Lambda$CDM point $(w_0,w_a)=(-1,0)$.}
\label{fig:w0wa}
\end{figure}

We also consider the $w_0w_a$CDM model, with the corresponding constraints presented in the second block of Table~\ref{table2}. As shown by the dashed magenta contours in Figure~\ref{fig:h0ob}(b), the resulting $H_0$--$\Omega_{\rm b}$ constraints are broader than those in $\Lambda$CDM due to the additional dark-energy degrees of freedom. Nevertheless, the preferred region remains consistent with the $\Lambda$CDM result and with the CMB, SH0ES, and BBN + CMB determinations, indicating that our main conclusions for $H_0$ and $\Omega_{\rm b}$ are robust to this model extension.

As for the dark-energy sector itself, Figure~\ref{fig:w0wa} shows that FRB + DESI alone leave a broad $w_0$-$w_a$ degeneracy, with the constraint dominated by the geometric information from BAO.
Adding SNe datasets sharply reduces this degeneracy. The FRB + DESI DR2 BAO + PantheonPlus combination gives
\begin{equation} \left. \begin{aligned} w_0 &= -0.887^{+0.057}_{-0.065},\\ w_a &= -0.22\pm0.44, \end{aligned} \right\} \begin{aligned} &{\rm FRB + DESI\ DR2\ BAO}\\ &+{\rm PantheonPlus}. \end{aligned} \end{equation}
Using 92 localized FRBs combined with DESI DR1 BAO and PantheonPlus data, \citet{Wang:2025ugc} obtained $w_{0}=-0.855\pm0.084$ and $w_{a}=-1.17^{+0.46}_{-0.49}$, reporting a $\sim2.5\sigma$ preference for dynamical dark energy. Our DESI DR2 result is more consistent with $\Lambda$CDM.
For DESY5, we obtain
$w_0=-0.780^{+0.066}_{-0.074}$ and $w_a=-0.76\pm0.45$,
with DES-Dovekie giving intermediate values.
The corresponding deviations from $\Lambda$CDM are $1.7\sigma$, $1.9\sigma$, and $3.2\sigma$ for PantheonPlus, DES-Dovekie, and DESY5, respectively, showing the dependence on the adopted SNe sample.
Thus, breaking the $\Omega_{\rm b}$--$H_0$ degeneracy does not remove the
$w_0$--$w_a$ degeneracy, and current dark-energy constraints remain dominated by BAO and SNe.

\subsection{Impact of DM-Modeling Systematics}
\label{sec:dm_systematics}

\begin{table}[htbp]
\caption{
Impact of FRB DM modeling on cosmological constraints from FRBs and DESI DR2 BAO in $\Lambda$CDM. The baseline adopts NE2001 for the Galactic ISM, a Gaussian halo DM contribution ($65\pm15\,\mathrm{pc\,cm^{-3}}$), fiducial IGM and host-DM distributions, and marginalizes over $f_{\rm d}$. Alternative cases modify the indicated Galactic foreground or extragalactic DM assumptions. TNG probability distribution functions (PDFs) are calibrated using TNG simulations, while ``fixed $f_{\rm d}$'' denotes $f_{\rm d}=0.94$. The sound horizon $r_{\rm d}$ is a derived parameter. All uncertainties represent $68\%$ credible intervals.
}
\label{table3}
\centering
\scriptsize
\setlength{\tabcolsep}{1.5pt}
\renewcommand{\arraystretch}{1.5}
\begin{tabular}{@{}lccc@{}}
\hline\hline
DM assumption & $H_0$ $[\mathrm{km\,s^{-1}\,Mpc^{-1}}]$ & $\Omega_{\rm b}$ & $r_{\rm d}$ $[\mathrm{Mpc}]$ \\
\hline
Baseline (FRB + DESI DR2) & $71.4^{+2.2}_{-2.5}$ & $0.0503^{+0.0025}_{-0.0021}$ & $142.4\pm4.8$ \\
\hline 
\multicolumn{4}{@{}l}{\textbf{\textit{Milky Way foreground}}} \\
YMW16 ISM & $71.6^{+2.2}_{-2.4}$ & $0.0505^{+0.0025}_{-0.0022}$ & $142.0\pm4.7$ \\
NE2025 ISM & $72.0\pm2.4$ & $0.0510^{+0.0026}_{-0.0021}$ & $141.1\pm4.8$ \\
YT2020 halo & $71.7\pm2.3$ & $0.0506^{+0.0025}_{-0.0021}$ & $141.9\pm4.7$ \\
Huang25 halo & $70.2\pm2.1$ & $0.0491^{+0.0025}_{-0.0020}$ & $144.9\pm4.4$ \\
\hline
\multicolumn{4}{@{}l}{\textbf{\textit{Extragalactic DM modeling}}} \\
Fixed $f_{\rm d}$ & $70.2^{+1.8}_{-2.0}$ & $0.0490^{+0.0022}_{-0.0018}$ & $144.8\pm 4.0$ \\
TNG PDFs & $69.9\pm1.5$ & $0.0488\pm0.0015$ & $145.1\pm3.2$ \\
TNG PDFs, fixed $f_{\rm d}$ & $68.8\pm1.0$ & $0.0475\pm0.0012$ & $147.7\pm2.5$ \\
\hline\hline
\end{tabular}
\end{table}

We assess the robustness of the FRB-BAO constraints against three classes of astrophysical uncertainty, including the Milky Way ISM contribution, the Milky Way halo contribution, and the probability distributions adopted for the IGM and host-galaxy DMs. The corresponding constraints are summarized in Table~\ref{table3}. In each test, we vary only the specified components while keeping all others fixed at their baseline choices.

\begin{figure}[htbp]
\centering
\includegraphics[width=0.48\textwidth]{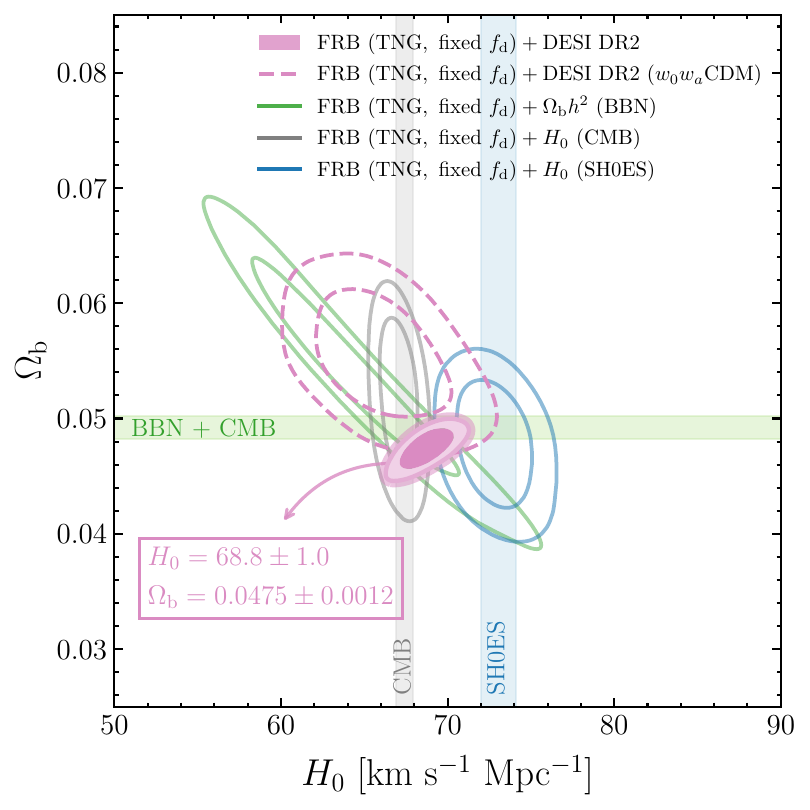}
\caption{Same as Figure~\ref{fig:h0ob}(b), but adopting TNG-calibrated FRB DM distributions and fixing the diffuse-baryon fraction to $f_{\rm d}=0.94$.}
\label{fig:tng}
\end{figure}

The inferred cosmological parameters are insensitive to the choice of Galactic foreground model at the current statistical precision. For example, replacing NE2001 with YMW16 gives $H_0=71.6^{+2.2}_{-2.4}~\mathrm{km\,s^{-1}~Mpc^{-1}}$, whereas adopting the Galactic-halo Huang25 model \citep{Huang:2024hgz} (with NE2001 retained for the Galactic ISM), gives $H_0=70.2\pm2.1~ \mathrm{km\,s^{-1}~Mpc^{-1}}$. Across all four alternative Galactic foreground models, even the largest shifts in $H_0$ and $\Omega_{\rm b}$ are only about their $0.5\sigma$ statistical uncertainties. The Huang25 model is the only case that shifts $H_0$ toward the CMB-inferred value, while the other three shift it slightly toward the local distance ladder. The tested Galactic foreground DM models therefore have only a modest impact on the current FRB-BAO constraints. Recent studies have also assessed Galactic foreground uncertainties and developed differential-DM methods to reduce their impact \citep{Xu:2025ddk,Zhang:2026lmm}.

A larger sensitivity arises from the modeling of the IGM and host-galaxy DM contributions.
Using TNG-calibrated IGM and host-galaxy DM distributions tightens the
joint constraints to
$H_0=69.9\pm1.5\,
\mathrm{km\,s^{-1}\,Mpc^{-1}}$ and
$\Omega_{\rm b}=0.0488\pm0.0015$.
Fixing $f_{\rm d}=0.94$ reported by \citet{Connor:2024mjg} further gives $H_0=68.8\pm1.0\,\mathrm{km\,s^{-1}\,Mpc^{-1}}$ and $\Omega_{\rm b}=0.0475\pm0.0012$, as shown in Figure~\ref{fig:tng}. Both cases shift $H_0$ toward the CMB-inferred value, while $\Omega_{\rm b}$ remains consistent with the BBN + CMB determination within $1.2\sigma$. This model sensitivity is also highlighted by recent studies of extragalactic DM distributions and their numerical or analytical modeling \citep{Konietzka:2025kdr,Jia:2025kvp,Zhuge:2025urk}. Under the TNG-calibrated DM model with $f_{\rm d}=0.94$ fixed, the forecast in Appendix~\ref{sec:forecast} reaches approximately $1.0\%$ statistical precision in $H_0$ with 500 localized FRBs.

Overall, these results demonstrate that our constraints are sensitive to the modeling of extragalactic DM. In the baseline analysis, which marginalizes over the extragalactic DM parameters, the inferred $H_0$ lies closer to the local distance-ladder value. In contrast, adopting TNG-calibrated DM distributions with fixed $f_{\rm d}$ shifts $H_0$ toward the CMB-inferred value, increasing the discrepancy with SH0ES to approximately $2.9\sigma$.

\section{Conclusion}\label{sec:conclusion}

In this work, we combine 120 localized FRBs with DESI DR2 BAO to simultaneously determine the Hubble constant $H_0$ and the cosmic baryon density $\Omega_{\rm b}$ from late-universe observations.
We find that the FRB and BAO constraints exhibit nearly orthogonal degeneracy directions in the $H_0$--$\Omega_{\rm b}$ plane, allowing their combination to efficiently break the individual parameter degeneracies. The joint analysis yields
$H_0=71.4^{+2.2}_{-2.5}\ {\rm km\ s^{-1}\ Mpc^{-1}}$
and $\Omega_{\rm b}=0.0503^{+0.0025}_{-0.0021}$ in $\Lambda$CDM,
corresponding to precisions of $3.3\%$ and $4.6\%$, respectively. 
The inferred $H_0$ is consistent with both the CMB and local distance-ladder determinations ($1.7\sigma$ and $0.6\sigma$, respectively), while $\Omega_{\rm b}$ agrees with the BBN + CMB constraint. These results demonstrate that the combination of FRBs and BAO can simultaneously constrain $H_0$ and $\Omega_{\rm b}$ without external CMB or BBN priors. 

We further assess the robustness of these measurements against FRB likelihood modeling. Different Galactic foreground models have only a modest impact at the current precision, whereas modeling the extragalactic DM distribution and diffuse-baryon fraction causes larger shifts and dominates the astrophysical modeling uncertainty. Adopting TNG-calibrated IGM and host-DM distributions with fixed $f_{\rm d}$ gives
$H_0=68.8\pm1.0\ {\rm km\,s^{-1}\,Mpc^{-1}}$ and
$\Omega_{\rm b}=0.0475\pm0.0012$,
shifting $H_0$ toward the CMB-inferred value. Under the same modeling assumptions, our forecasts indicate that 500 localized FRBs could achieve approximately $1.0\%$ precision in $H_0$.

Overall, our results demonstrate the strong complementarity of FRB and DESI BAO for jointly constraining cosmic expansion and baryon abundance. Larger localized FRB samples and more precise BAO measurements can further tighten these constraints, provided that modeling of the IGM, host-galaxy DM, and diffuse-baryon fraction improves accordingly. The FRB-BAO combination therefore offers a distinct route to increasingly precise and robust late-universe measurements of both $H_{0}$ and $\Omega_{\rm b}$, providing an independent consistency test of the standard cosmological model.

\section*{Acknowledgments}

We thank Ze-Wei Zhao for fruitful discussions. This work was supported by the National Natural Science Foundation of China (Grants Nos. 12533001, 12575049, and 12473001), the National SKA Program of China (Grants Nos. 2022SKA0110200 and 2022SKA0110203), the China Manned Space Program (Grant No. CMS-CSST-2025-A02), and the National 111 Project (Grant No. B16009).

\bibliography{main}

\begin{thebibliography}{}
\expandafter\ifx\csname natexlab\endcsname\relax\def\natexlab#1{#1}\fi
\providecommand{\url}[1]{\href{#1}{#1}}
\providecommand{\dodoi}[1]{doi:~\href{http://doi.org/#1}{\nolinkurl{#1}}}
\providecommand{\doeprint}[1]{\href{http://ascl.net/#1}{\nolinkurl{http://ascl.net/#1}}}
\providecommand{\doarXiv}[1]{\href{https://arxiv.org/abs/#1}{\nolinkurl{https://arxiv.org/abs/#1}}}

\bibitem[{Abbott {et~al.}(2024)}]{DES:2024jxu}
Abbott, T. M.~C., {et~al.} 2024, Astrophys. J. Lett., 973, L14, \dodoi{10.3847/2041-8213/ad6f9f}

\bibitem[{Abdalla {et~al.}(2022)}]{Abdalla:2022yfr}
Abdalla, E., {et~al.} 2022, JHEAp, 34, 49, \dodoi{10.1016/j.jheap.2022.04.002}

\bibitem[{Abdul~Karim {et~al.}(2025{\natexlab{a}})}]{DESI:2025zgx}
Abdul~Karim, M., {et~al.} 2025{\natexlab{a}}, Phys. Rev. D, 112, 083515, \dodoi{10.1103/tr6y-kpc6}

\bibitem[{Abdul~Karim {et~al.}(2025{\natexlab{b}})}]{DESI:2025zpo}
---. 2025{\natexlab{b}}, Phys. Rev. D, 112, 083514, \dodoi{10.1103/2wwn-xjm5}

\bibitem[{Acharya \& Beniamini(2025)}]{Acharya:2025ubt}
Acharya, S.~K., \& Beniamini, P. 2025, JCAP, 10, 073, \dodoi{10.1088/1475-7516/2025/10/073}

\bibitem[{Aghanim {et~al.}(2020)}]{Planck:2018vyg}
Aghanim, N., {et~al.} 2020, Astron. Astrophys., 641, A6, \dodoi{10.1051/0004-6361/201833910}

\bibitem[{Amiri {et~al.}(2025)}]{CHIMEFRB:2025ggb}
Amiri, M., {et~al.} 2025, Astrophys. J. Suppl., 280, 6, \dodoi{10.3847/1538-4365/addbda}

\bibitem[{Anna-Thomas {et~al.}(2025)}]{Anna-Thomas:2025zao}
Anna-Thomas, R., {et~al.} 2025, Astrophys. J., 993, 221, \dodoi{10.3847/1538-4357/ae1014}

\bibitem[{Bannister {et~al.}(2019)}]{Bannister:2019iju}
Bannister, K.~W., {et~al.} 2019, \dodoi{10.1126/science.aaw5903}

\bibitem[{Bella {et~al.}(2026)Bella, Poulin, Vagnozzi, \& Knox}]{Bella:2026zuk}
Bella, M., Poulin, V., Vagnozzi, S., \& Knox, L. 2026.
\newblock \doarXiv{2604.13535}

\bibitem[{Bhandari {et~al.}(2020)}]{Bhandari:2020oyb}
Bhandari, S., {et~al.} 2020, Astrophys. J. Lett., 895, L37, \dodoi{10.3847/2041-8213/ab672e}

\bibitem[{Bhandari {et~al.}(2022)}]{Bhandari:2021pvj}
---. 2022, Astron. J., 163, 69, \dodoi{10.3847/1538-3881/ac3aec}

\bibitem[{Bhandari {et~al.}(2023)}]{Bhandari:2022ton}
---. 2023, Astrophys. J., 948, 67, \dodoi{10.3847/1538-4357/acc178}

\bibitem[{Bhardwaj {et~al.}(2024)}]{Bhardwaj:2023vha}
Bhardwaj, M., {et~al.} 2024, Astrophys. J. Lett., 971, L51, \dodoi{10.3847/2041-8213/ad64d1}

\bibitem[{Birrer {et~al.}(2025)}]{TDCOSMO:2025dmr}
Birrer, S., {et~al.} 2025, Astron. Astrophys., 704, A63, \dodoi{10.1051/0004-6361/202555801}

\bibitem[{Brout {et~al.}(2022)}]{Brout:2022vxf}
Brout, D., {et~al.} 2022, Astrophys. J., 938, 110, \dodoi{10.3847/1538-4357/ac8e04}

\bibitem[{Cai \& Wang(2026)}]{Cai:2026swf}
Cai, R.-G., \& Wang, S.-J. 2026, Res. Astron. Astrophys., 26, 084011, \dodoi{10.1088/1674-4527/ae842f}

\bibitem[{Cai {et~al.}(2026)Cai, Ren, Qiu, Li, \& Zhang}]{Cai:2025mas}
Cai, Y., Ren, X., Qiu, T., Li, M., \& Zhang, X. 2026, Natl. Sci. Rev., 13, nwag115, \dodoi{10.1093/nsr/nwag115}

\bibitem[{Caleb {et~al.}(2023)}]{Caleb:2023atr}
Caleb, M., {et~al.} 2023, Mon. Not. Roy. Astron. Soc., 524, 2064, \dodoi{10.1093/mnras/stad1839}

\bibitem[{Caleb {et~al.}(2025)}]{Caleb:2025uzd}
---. 2025.
\newblock \doarXiv{2508.01648}

\bibitem[{Cassanelli {et~al.}(2024)}]{Cassanelli:2023hvg}
Cassanelli, T., {et~al.} 2024, Nature Astron., 8, 1429, \dodoi{10.1038/s41550-024-02357-x}

\bibitem[{Cen \& Ostriker(1999)}]{Cen:1998hc}
Cen, R., \& Ostriker, J.~P. 1999, Astrophys. J., 514, 1, \dodoi{10.1086/306949}

\bibitem[{Chatterjee {et~al.}(2017)}]{Chatterjee:2017dqg}
Chatterjee, S., {et~al.} 2017, Nature, 541, 58, \dodoi{10.1038/nature20797}

\bibitem[{Chevallier \& Polarski(2001)}]{Chevallier:2000qy}
Chevallier, M., \& Polarski, D. 2001, Int. J. Mod. Phys. D, 10, 213, \dodoi{10.1142/S0218271801000822}

\bibitem[{Chittidi {et~al.}(2021)Chittidi, Simha, Mannings, Prochaska, Ryder, Rafelski, Neeleman, Macquart, Tejos, Jorgenson, Day, Marnoch, Bhandari, Deller, Qiu, Bannister, Shannon, \& Heintz}]{Chittidi2021}
Chittidi, J.~S., Simha, S., Mannings, A., {et~al.} 2021, Astrophys. J., 922, 173, \dodoi{10.3847/1538-4357/ac2818}

\bibitem[{Connor {et~al.}(2025)}]{Connor:2024mjg}
Connor, L., {et~al.} 2025, Nature Astron., 9, 1226, \dodoi{10.1038/s41550-025-02566-y}

\bibitem[{Cordes \& Lazio(2002)}]{Cordes:2002wz}
Cordes, J.~M., \& Lazio, T. J.~W. 2002.
\newblock \doarXiv{astro-ph/0207156}

\bibitem[{Curtin {et~al.}(2026)}]{Curtin:2025tvg}
Curtin, A.~P., {et~al.} 2026, Astrophys. J., 998, 97, \dodoi{10.3847/1538-4357/ae2ea0}

\bibitem[{Dai(2023)}]{Dai:2023ska}
Dai, Z.-G. 2023, Science China. Physics, Mechanics \& Astronomy, 66, 120431

\bibitem[{Deng \& Zhang(2014)}]{Deng:2013aga}
Deng, W., \& Zhang, B. 2014, Astrophys. J. Lett., 783, L35, \dodoi{10.1088/2041-8205/783/2/L35}

\bibitem[{Di~Valentino {et~al.}(2020)Di~Valentino, Melchiorri, Mena, \& Vagnozzi}]{DiValentino:2019ffd}
Di~Valentino, E., Melchiorri, A., Mena, O., \& Vagnozzi, S. 2020, Phys. Dark Univ., 30, 100666, \dodoi{10.1016/j.dark.2020.100666}

\bibitem[{Di~Valentino {et~al.}(2021{\natexlab{a}})}]{DiValentino:2020zio}
Di~Valentino, E., {et~al.} 2021{\natexlab{a}}, Astropart. Phys., 131, 102605, \dodoi{10.1016/j.astropartphys.2021.102605}

\bibitem[{Di~Valentino {et~al.}(2021{\natexlab{b}})Di~Valentino, Mena, Pan, Visinelli, Yang, Melchiorri, Mota, Riess, \& Silk}]{DiValentino:2021izs}
Di~Valentino, E., Mena, O., Pan, S., {et~al.} 2021{\natexlab{b}}, Class. Quant. Grav., 38, 153001, \dodoi{10.1088/1361-6382/ac086d}

\bibitem[{Du {et~al.}(2026{\natexlab{a}})Du, Li, Wu, Zhang, \& Zhang}]{Du:2025xes}
Du, G.-H., Li, T.-N., Wu, P.-J., Zhang, J.-F., \& Zhang, X. 2026{\natexlab{a}}, Sci. China Phys. Mech. Astron., 69, 100413, \dodoi{10.1007/s11433-026-3051-4}

\bibitem[{Du {et~al.}(2026{\natexlab{b}})Du, Li, Yin, Zhou, Wang, Zhang, \& Zhang}]{Du:2026qtq}
Du, G.-H., Li, T.-N., Yin, L., {et~al.} 2026{\natexlab{b}}, Astrophys. J. Lett., 1008, L62, \dodoi{10.3847/2041-8213/ae9f47}

\bibitem[{Eftekhari {et~al.}(2025)}]{Eftekhari:2024bmi}
Eftekhari, T., {et~al.} 2025, Astrophys. J. Lett., 979, L22, \dodoi{10.3847/2041-8213/ad9de2}

\bibitem[{Fortunato {et~al.}(2025)Fortunato, Bacon, Hip{\'o}lito-Ricaldi, \& Wands}]{Fortunato:2024hfm}
Fortunato, J. A.~S., Bacon, D.~J., Hip{\'o}lito-Ricaldi, W.~S., \& Wands, D. 2025, JCAP, 01, 018, \dodoi{10.1088/1475-7516/2025/01/018}

\bibitem[{Fortunato {et~al.}(2026)Fortunato, Hip{\'o}lito-Ricaldi, \& Romero}]{Fortunato:2025qxc}
Fortunato, J. A.~S., Hip{\'o}lito-Ricaldi, W.~S., \& Romero, G.~E. 2026, Phys. Rev. D, 113, 023537, \dodoi{10.1103/4vgl-j1r7}

\bibitem[{Fukugita {et~al.}(1998)Fukugita, Hogan, \& Peebles}]{Fukugita:1997bi}
Fukugita, M., Hogan, C.~J., \& Peebles, P. J.~E. 1998, Astrophys. J., 503, 518, \dodoi{10.1086/306025}

\bibitem[{Gao {et~al.}(2025)Gao, Wu, Hu, Yi, Zhou, Wang, \& Dai}]{Gao:2025fcr}
Gao, D.~H., Wu, Q., Hu, J.~P., {et~al.} 2025, Astron. Astrophys., 698, A215, \dodoi{10.1051/0004-6361/202453006}

\bibitem[{Gao {et~al.}(2014)Gao, Li, \& Zhang}]{Gao:2014iva}
Gao, H., Li, Z., \& Zhang, B. 2014, Astrophys. J., 788, 189, \dodoi{10.1088/0004-637X/788/2/189}

\bibitem[{Gao {et~al.}(2023)Gao, Zhou, Du, Zou, Hu, \& Xu}]{Gao:2023izj}
Gao, J., Zhou, Z., Du, M., {et~al.} 2023, Mon. Not. Roy. Astron. Soc., 527, 7861, \dodoi{10.1093/mnras/stad3708}

\bibitem[{Gelman \& Rubin(1992)}]{Gelman:1992zz}
Gelman, A., \& Rubin, D.~B. 1992, Statist. Sci., 7, 457, \dodoi{10.1214/ss/1177011136}

\bibitem[{Giar{\`e}(2023)}]{Giare:2023xoc}
Giar{\`e}, W. 2023, \dodoi{10.1007/978-981-99-0177-7\_36}

\bibitem[{Giar{\`e} {et~al.}(2026)Giar{\`e}, Lee, \& Di~Valentino}]{Giare:2026oti}
Giar{\`e}, W., Lee, D.~H., \& Di~Valentino, E. 2026, New Astron. Rev., 103, 101759, \dodoi{10.1016/j.newar.2026.101759}

\bibitem[{Giar{\`e} {et~al.}(2024{\natexlab{a}})Giar{\`e}, Najafi, Pan, Di~Valentino, \& Firouzjaee}]{Giare:2024gpk}
Giar{\`e}, W., Najafi, M., Pan, S., Di~Valentino, E., \& Firouzjaee, J.~T. 2024{\natexlab{a}}, JCAP, 10, 035, \dodoi{10.1088/1475-7516/2024/10/035}

\bibitem[{Giar{\`e} {et~al.}(2024{\natexlab{b}})Giar{\`e}, Sabogal, Nunes, \& Di~Valentino}]{Giare:2024smz}
Giar{\`e}, W., Sabogal, M.~A., Nunes, R.~C., \& Di~Valentino, E. 2024{\natexlab{b}}, Phys. Rev. Lett., 133, 251003, \dodoi{10.1103/PhysRevLett.133.251003}

\bibitem[{Gordon {et~al.}(2023)}]{Gordon:2023cgw}
Gordon, A.~C., {et~al.} 2023, Astrophys. J., 954, 80, \dodoi{10.3847/1538-4357/ace5aa}

\bibitem[{Gordon {et~al.}(2025)}]{Gordon:2025spatial}
---. 2025.
\newblock \doarXiv{2506.06453}

\bibitem[{Guo {et~al.}(2019)Guo, Zhang, \& Zhang}]{Guo:2018ans}
Guo, R.-Y., Zhang, J.-F., \& Zhang, X. 2019, JCAP, 02, 054, \dodoi{10.1088/1475-7516/2019/02/054}

\bibitem[{Hagstotz {et~al.}(2022)Hagstotz, Reischke, \& Lilow}]{Hagstotz:2021jzu}
Hagstotz, S., Reischke, R., \& Lilow, R. 2022, Mon. Not. Roy. Astron. Soc., 511, 662, \dodoi{10.1093/mnras/stac077}

\bibitem[{Hanmer {et~al.}(2025)}]{Hanmer:2025msu}
Hanmer, K.~Y., {et~al.} 2025, Mon. Not. Roy. Astron. Soc., 538, 1800, \dodoi{10.1093/mnras/staf289}

\bibitem[{Heintz {et~al.}(2020)Heintz, Prochaska, Simha, Platts, fai Fong, Tejos, Ryder, Aggarwal, Bhandari, Day, Deller, Kilpatrick, Law, Macquart, Mannings, Marnoch, Sadler, \& Shannon}]{Heintz2020}
Heintz, K.~E., Prochaska, J.~X., Simha, S., {et~al.} 2020, Astrophys. J., 903, 152, \dodoi{10.3847/1538-4357/abb6fb}

\bibitem[{Hu \& Wang(2023)}]{Hu:2023jqc}
Hu, J.-P., \& Wang, F.-Y. 2023, Universe, 9, 94, \dodoi{10.3390/universe9020094}

\bibitem[{Huang {et~al.}(2025)Huang, Lee, Libeskind, Simha, Valade, \& Prochaska}]{Huang:2024hgz}
Huang, Y., Lee, K.-G., Libeskind, N.~I., {et~al.} 2025, Mon. Not. Roy. Astron. Soc., 538, 2785, \dodoi{10.1093/mnras/staf417}

\bibitem[{Ibik {et~al.}(2024)}]{Ibik:2023ugl}
Ibik, A.~L., {et~al.} 2024, Astrophys. J., 961, 99, \dodoi{10.3847/1538-4357/ad0893}

\bibitem[{Inoue(2004)}]{Inoue:2003ga}
Inoue, S. 2004, Mon. Not. Roy. Astron. Soc., 348, 999, \dodoi{10.1111/j.1365-2966.2004.07359.x}

\bibitem[{Ioka(2003)}]{Ioka:2003fr}
Ioka, K. 2003, Astrophys. J. Lett., 598, L79, \dodoi{10.1086/380598}

\bibitem[{Jahns-Schindler \& Spitler(2025)}]{Jahns-Schindler:2025saf}
Jahns-Schindler, J.~N., \& Spitler, L.~G. 2025, Phys. Rev. D, 112, 103541, \dodoi{10.1103/wtv1-dmrw}

\bibitem[{James {et~al.}(2022)}]{James:2022dcx}
James, C.~W., {et~al.} 2022, Mon. Not. Roy. Astron. Soc., 516, 4862, \dodoi{10.1093/mnras/stac2524}

\bibitem[{Jia {et~al.}(2026)Jia, Qiang, Li, \& Wei}]{Jia:2025kvp}
Jia, J.-Y., Qiang, D.-C., Li, L.-Y., \& Wei, H. 2026, Phys. Lett. B, 877, 140473, \dodoi{10.1016/j.physletb.2026.140473}

\bibitem[{Jiang {et~al.}(2024)Jiang, Pedrotti, da~Costa, \& Vagnozzi}]{Jiang:2024xnu}
Jiang, J.-Q., Pedrotti, D., da~Costa, S.~S., \& Vagnozzi, S. 2024, Phys. Rev. D, 110, 123519, \dodoi{10.1103/PhysRevD.110.123519}

\bibitem[{Kalita {et~al.}(2025)Kalita, Bhatporia, \& Weltman}]{Kalita:2024xae}
Kalita, S., Bhatporia, S., \& Weltman, A. 2025, Phys. Dark Univ., 48, 101926, \dodoi{10.1016/j.dark.2025.101926}

\bibitem[{Khrykin {et~al.}(2024)}]{Khrykin:2024lmd}
Khrykin, I.~S., {et~al.} 2024, Astrophys. J., 973, 151, \dodoi{10.3847/1538-4357/ad6567}

\bibitem[{Konietzka {et~al.}(2025)Konietzka, Connor, Semenov, Beane, Springel, \& Hernquist}]{Konietzka:2025kdr}
Konietzka, R.~M., Connor, L., Semenov, V.~A., {et~al.} 2025.
\newblock \doarXiv{2507.07090}

\bibitem[{Kumar \& Linder(2019)}]{Kumar:2019qhc}
Kumar, P., \& Linder, E.~V. 2019, Phys. Rev. D, 100, 083533, \dodoi{10.1103/PhysRevD.100.083533}

\bibitem[{Law {et~al.}(2020)}]{Law:2020cnm}
Law, C.~J., {et~al.} 2020, Astrophys. J., 899, 161, \dodoi{10.3847/1538-4357/aba4ac}

\bibitem[{Law {et~al.}(2024)}]{Law:2023ibd}
---. 2024, Astrophys. J., 967, 29, \dodoi{10.3847/1538-4357/ad3736}

\bibitem[{Lemos(2025)}]{Lemos:2025bgy}
Lemos, T. 2025, JCAP, 11, 060, \dodoi{10.1088/1475-7516/2025/11/060}

\bibitem[{Leung {et~al.}(2026)Leung, Borrow, Masui, Andrew, Chen, Schaye, \& Schaller}]{Leung:2025vej}
Leung, C., Borrow, J., Masui, K.~W., {et~al.} 2026, Phys. Rev. Lett., 137, 131003, \dodoi{10.1103/ms7q-gw8d}

\bibitem[{Lewis(2025)}]{Lewis:2019xzd}
Lewis, A. 2025, JCAP, 08, 025, \dodoi{10.1088/1475-7516/2025/08/025}

\bibitem[{Li \& Wang(2025)}]{Li:2025ops}
Li, J.-X., \& Wang, S. 2025, Eur. Phys. J. C, 85, 1308, \dodoi{10.1140/epjc/s10052-025-15065-1}

\bibitem[{Li {et~al.}(2026{\natexlab{a}})Li, Du, Li, Wu, Jin, Zhang, \& Zhang}]{Li:2025owk}
Li, T.-N., Du, G.-H., Li, Y.-H., {et~al.} 2026{\natexlab{a}}, Sci. China Phys. Mech. Astron., 69, 210413, \dodoi{10.1007/s11433-025-2771-5}

\bibitem[{Li {et~al.}(2026{\natexlab{b}})Li, Du, Wang, Li, Zhang, \& Zhang}]{Li:2026asg}
Li, T.-N., Du, G.-H., Wang, H., {et~al.} 2026{\natexlab{b}}, Res. Astron. Astrophys., 26, 084002, \dodoi{10.1088/1674-4527/ae8429}

\bibitem[{Li {et~al.}(2026{\natexlab{c}})Li, Du, Zhou, Li, Zhang, \& Zhang}]{Li:2025vuh}
Li, T.-N., Du, G.-H., Zhou, S.-H., {et~al.} 2026{\natexlab{c}}, Phys. Dark Univ., 52, 102254, \dodoi{10.1016/j.dark.2026.102254}

\bibitem[{Li {et~al.}(2026{\natexlab{d}})Li, Giar{\`e}, Du, Li, Di~Valentino, Zhang, \& Zhang}]{Li:2026xaz}
Li, T.-N., Giar{\`e}, W., Du, G.-H., {et~al.} 2026{\natexlab{d}}.
\newblock \doarXiv{2601.07361}

\bibitem[{Li {et~al.}(2026{\natexlab{e}})Li, Liu, Li, Du, Shafieloo, \& Biesiada}]{Li:2026hwq}
Li, X., Liu, T., Li, T.-N., {et~al.} 2026{\natexlab{e}}, Astrophys. J. Lett., 1001, L21, \dodoi{10.3847/2041-8213/ae5a36}

\bibitem[{Li {et~al.}(2026{\natexlab{f}})}]{Li:2025ckl}
Li, Y., {et~al.} 2026{\natexlab{f}}, Science, 391, 280, \dodoi{10.1126/science.adq3225}

\bibitem[{Li {et~al.}(2019)Li, Gao, Wei, Yang, Zhang, \& Zhu}]{Li:2019klc}
Li, Z., Gao, H., Wei, J.-J., {et~al.} 2019, Astrophys. J., 876, 146, \dodoi{10.3847/1538-4357/ab18fe}

\bibitem[{Li {et~al.}(2020)Li, Gao, Wei, Yang, Zhang, \& Zhu}]{Li:2020qei}
---. 2020, Mon. Not. Roy. Astron. Soc., 496, L28, \dodoi{10.1093/mnrasl/slaa070}

\bibitem[{Linder(2003)}]{Linder:2002et}
Linder, E.~V. 2003, Phys. Rev. Lett., 90, 091301, \dodoi{10.1103/PhysRevLett.90.091301}

\bibitem[{Liu {et~al.}(2025)Liu, Li, \& Wang}]{Liu:2025mub}
Liu, T., Li, X., \& Wang, J. 2025, Astrophys. J., 988, 243, \dodoi{10.3847/1538-4357/aded8b}

\bibitem[{Liu {et~al.}(2026{\natexlab{a}})Liu, Wang, Wu, Wei, \& Wu}]{Liu:2026txt}
Liu, Y., Wang, B., Wu, P., Wei, J.-J., \& Wu, X.-F. 2026{\natexlab{a}}, Astrophys. J., 998, 15, \dodoi{10.3847/1538-4357/ae355a}

\bibitem[{Liu {et~al.}(2026{\natexlab{b}})Liu, Wei, Wu, \& Wu}]{Liu:2026epv}
Liu, Y., Wei, J.-J., Wu, P., \& Wu, X.-F. 2026{\natexlab{b}}.
\newblock \doarXiv{2604.03769}

\bibitem[{Liu {et~al.}(2023)Liu, Yu, \& Wu}]{Liu:2022bmn}
Liu, Y., Yu, H., \& Wu, P. 2023, Astrophys. J. Lett., 946, L49, \dodoi{10.3847/2041-8213/acc650}

\bibitem[{Liu {et~al.}(2026{\natexlab{c}})Liu, Zhang, Wei, Wu, Yu, \& Wu}]{Liu:2025fdf}
Liu, Y., Zhang, Y., Wei, J.-J., {et~al.} 2026{\natexlab{c}}, Phys. Rev. D, 114, 043537, \dodoi{10.1103/mdvc-wq7t}

\bibitem[{Lorimer {et~al.}(2007)Lorimer, Bailes, McLaughlin, Narkevic, \& Crawford}]{Lorimer:2007qn}
Lorimer, D.~R., Bailes, M., McLaughlin, M.~A., Narkevic, D.~J., \& Crawford, F. 2007, Science, 318, 777, \dodoi{10.1126/science.1147532}

\bibitem[{Macquart {et~al.}(2020)}]{Macquart:2020lln}
Macquart, J.~P., {et~al.} 2020, Nature, 581, 391, \dodoi{10.1038/s41586-020-2300-2}

\bibitem[{McQuinn(2014)}]{McQuinn:2013tmc}
McQuinn, M. 2014, Astrophys. J. Lett., 780, L33, \dodoi{10.1088/2041-8205/780/2/L33}

\bibitem[{Michilli {et~al.}(2023)}]{Michilli:2022bbs}
Michilli, D., {et~al.} 2023, Astrophys. J., 950, 134, \dodoi{10.3847/1538-4357/accf89}

\bibitem[{Mukherjee {et~al.}(2026)Mukherjee, Shafieloo, Sahni, \& Shtanov}]{Mukherjee:2026lnb}
Mukherjee, P., Shafieloo, A., Sahni, V., \& Shtanov, Y. 2026.
\newblock \doarXiv{2609.39039}

\bibitem[{Nelson {et~al.}(2019)}]{Nelson:2018uso}
Nelson, D., {et~al.} 2019, Comput. Astrophys. Cosmol., 6, 2, \dodoi{10.1186/s40668-019-0028-x}

\bibitem[{Ocker \& Cordes(2026)}]{Ocker:2026mta}
Ocker, S.~K., \& Cordes, J.~M. 2026, Astrophys. J., 1002, 3, \dodoi{10.3847/1538-4357/ae5825}

\bibitem[{Pan {et~al.}(2026)Pan, Paul, Saridakis, \& Yang}]{Pan:2025qwy}
Pan, S., Paul, S., Saridakis, E.~N., \& Yang, W. 2026, Phys. Rev. D, 113, 023515, \dodoi{10.1103/5y21-k39n}

\bibitem[{Pang {et~al.}(2025)Pang, Zhang, \& Huang}]{Pang:2025lvh}
Pang, Y.-H., Zhang, X., \& Huang, Q.-G. 2025, Sci. China Phys. Mech. Astron., 68, 280410, \dodoi{10.1007/s11433-025-2713-8}

\bibitem[{Pastor-Marazuela {et~al.}(2026)}]{Pastor-Marazuela:2025loc}
Pastor-Marazuela, I., {et~al.} 2026, Mon. Not. Roy. Astron. Soc., 545, staf2144, \dodoi{10.1093/mnras/staf2144}

\bibitem[{Pedrotti {et~al.}(2026)Pedrotti, Giar{\`e}, Cheng, \& Di~Valentino}]{Pedrotti:2026ncd}
Pedrotti, D., Giar{\`e}, W., Cheng, H., \& Di~Valentino, E. 2026.
\newblock \doarXiv{2610.08728}

\bibitem[{Pillepich {et~al.}(2018)}]{Pillepich:2017jle}
Pillepich, A., {et~al.} 2018, Mon. Not. Roy. Astron. Soc., 473, 4077, \dodoi{10.1093/mnras/stx2656}

\bibitem[{Piratova-Moreno {et~al.}(2025)Piratova-Moreno, Garc{\'\i}a, Benavides-Gallego, \& Cabrera}]{Piratova-Moreno:2025cpc}
Piratova-Moreno, E.~F., Garc{\'\i}a, L.~{\'A}., Benavides-Gallego, C.~A., \& Cabrera, C. 2025.
\newblock \doarXiv{2502.08509}

\bibitem[{Popovic {et~al.}(2026)}]{DES:2025sig}
Popovic, B., {et~al.} 2026, Mon. Not. Roy. Astron. Soc., 548, stag632, \dodoi{10.1093/mnras/stag632}

\bibitem[{Poulin {et~al.}(2023)Poulin, Smith, \& Karwal}]{Poulin:2023lkg}
Poulin, V., Smith, T.~L., \& Karwal, T. 2023, Phys. Dark Univ., 42, 101348, \dodoi{10.1016/j.dark.2023.101348}

\bibitem[{Price {et~al.}(2021)Price, Flynn, \& Deller}]{Price:2021gzo}
Price, D.~C., Flynn, C., \& Deller, A. 2021, Publ. Astron. Soc. Austral., 38, e038, \dodoi{10.1017/pasa.2021.33}

\bibitem[{Prochaska \& Zheng(2019)}]{Prochaska_2019}
Prochaska, J.~X., \& Zheng, Y. 2019, Mon. Not. Roy. Astron. Soc., 485, 648, \dodoi{10.1093/mnras/stz261}

\bibitem[{Prochaska {et~al.}(2019)Prochaska, Macquart, McQuinn, Simha, Shannon, Day, Marnoch, Ryder, Deller, Bannister, Bhandari, Bordoloi, Bunton, Cho, Flynn, Mahony, Phillips, Qiu, \& Tejos}]{Prochaska:2019mnv}
Prochaska, J.~X., Macquart, J.-P., McQuinn, M., {et~al.} 2019, Science, 366, 231, \dodoi{10.1126/science.aay0073}

\bibitem[{Qiu {et~al.}(2022)Qiu, Zhao, Wang, Zhang, \& Zhang}]{Qiu:2021cww}
Qiu, X.-W., Zhao, Z.-W., Wang, L.-F., Zhang, J.-F., \& Zhang, X. 2022, JCAP, 02, 006, \dodoi{10.1088/1475-7516/2022/02/006}

\bibitem[{Rajwade {et~al.}(2024)}]{Rajwade:2024ozu}
Rajwade, K.~M., {et~al.} 2024, Mon. Not. Roy. Astron. Soc., 532, 3881, \dodoi{10.1093/mnras/stae1652}

\bibitem[{Ravi {et~al.}(2019)}]{Ravi:2019alc}
Ravi, V., {et~al.} 2019, Nature, 572, 352, \dodoi{10.1038/s41586-019-1389-7}

\bibitem[{Ravi {et~al.}(2022)}]{Ravi:2021kqk}
---. 2022, Mon. Not. Roy. Astron. Soc., 513, 982, \dodoi{10.1093/mnras/stac465}

\bibitem[{Ren {et~al.}(2026)Ren, Li, Liu, Cai, Li, \& Zhang}]{Ren:2026jyw}
Ren, X., Li, S.-Y., Liu, Y., {et~al.} 2026, Res. Astron. Astrophys., 26, 084001, \dodoi{10.1088/1674-4527/ae842b}

\bibitem[{Ribeiro {et~al.}(2026)Ribeiro, Sales, de~Farias, Santos, Batista, \& Queiroz}]{Ribeiro:2026lyd}
Ribeiro, B. W.~N., Sales, L.~L., de~Farias, K. E.~L., {et~al.} 2026.
\newblock \doarXiv{2606.29583}

\bibitem[{Riess {et~al.}(2022)}]{Riess:2021jrx}
Riess, A.~G., {et~al.} 2022, Astrophys. J. Lett., 934, L7, \dodoi{10.3847/2041-8213/ac5c5b}

\bibitem[{Ryder {et~al.}(2023)}]{Ryder:2022qpg}
Ryder, S.~D., {et~al.} 2023, Science, 392, 294, \dodoi{10.1126/science.adf2678}

\bibitem[{Sabogal {et~al.}(2026)Sabogal, Escamilla, Pedrotti, Selimaj, \& Vagnozzi}]{Sabogal:2026ipu}
Sabogal, M.~A., Escamilla, L.~A., Pedrotti, D., Selimaj, E., \& Vagnozzi, S. 2026.
\newblock \doarXiv{2607.21244}

\bibitem[{Sales {et~al.}(2026)Sales, de~Farias, Queiroz, Batista, Ribeiro, \& Santos}]{Sales:2026smc}
Sales, L.~L., de~Farias, K. E.~L., Queiroz, A.~R., {et~al.} 2026.
\newblock \doarXiv{2605.01990}

\bibitem[{Sch{\"o}neberg(2024)}]{Schoneberg:2024ifp}
Sch{\"o}neberg, N. 2024, JCAP, 06, 006, \dodoi{10.1088/1475-7516/2024/06/006}

\bibitem[{Sch{\"o}neberg {et~al.}(2022)Sch{\"o}neberg, Franco~Abell{\'a}n, P{\'e}rez~S{\'a}nchez, Witte, Poulin, \& Lesgourgues}]{Schoneberg:2021qvd}
Sch{\"o}neberg, N., Franco~Abell{\'a}n, G., P{\'e}rez~S{\'a}nchez, A., {et~al.} 2022, Phys. Rept., 984, 1, \dodoi{10.1016/j.physrep.2022.07.001}

\bibitem[{Sch{\"o}neberg {et~al.}(2026)Sch{\"o}neberg, Poulin, Ferrari, Finelli, Lesgourgues, Morelli, Mosbech, Sharma, \& Simon}]{Schoneberg:2026vaf}
Sch{\"o}neberg, N., Poulin, V., Ferrari, A.~G., {et~al.} 2026.
\newblock \doarXiv{2607.13283}

\bibitem[{Scolnic {et~al.}(2022)}]{Scolnic:2021amr}
Scolnic, D., {et~al.} 2022, Astrophys. J., 938, 113, \dodoi{10.3847/1538-4357/ac8b7a}

\bibitem[{Shannon {et~al.}(2025)}]{Shannon:2024pbu}
Shannon, R.~M., {et~al.} 2025, Publ. Astron. Soc. Austral., 42, e036, \dodoi{10.1017/pasa.2025.8}

\bibitem[{Sharma {et~al.}(2026{\natexlab{a}})Sharma, Krause, Ravi, Connor, Anbajagane, \& Rajendra~Singh}]{Sharma:2026xbj}
Sharma, K., Krause, E., Ravi, V., {et~al.} 2026{\natexlab{a}}, Nature Astron., \dodoi{10.1038/s41550-026-02957-9}

\bibitem[{Sharma {et~al.}(2026{\natexlab{b}})Sharma, Ravi, Connor, Krause, S., \& Anbajagane}]{Sharma:2025lql}
Sharma, K., Ravi, V., Connor, L., {et~al.} 2026{\natexlab{b}}, Astrophys. J., 999, 202, \dodoi{10.3847/1538-4357/ae4696}

\bibitem[{Sharma {et~al.}(2024)}]{Sharma:2024fsq}
Sharma, K., {et~al.} 2024, Nature, 635, 61, \dodoi{10.1038/s41586-024-08074-9}

\bibitem[{Shin {et~al.}(2025)}]{Shin:2024ezg}
Shin, K., {et~al.} 2025, Astrophys. J., 993, 208, \dodoi{10.3847/1538-4357/ae093b}

\bibitem[{Shull {et~al.}(2012)Shull, Smith, \& Danforth}]{Shull:2011aa}
Shull, J.~M., Smith, B.~D., \& Danforth, C.~W. 2012, Astrophys. J., 759, 23, \dodoi{10.1088/0004-637X/759/1/23}

\bibitem[{Tendulkar {et~al.}(2017)}]{Tendulkar:2017vuq}
Tendulkar, S.~P., {et~al.} 2017, Astrophys. J. Lett., 834, L7, \dodoi{10.3847/2041-8213/834/2/L7}

\bibitem[{Thornton {et~al.}(2013)}]{Thornton:2013iua}
Thornton, D., {et~al.} 2013, Science, 341, 53, \dodoi{10.1126/science.1236789}

\bibitem[{Tian {et~al.}(2024)Tian, Rajwade, Pastor-Marazuela, Stappers, Bezuidenhout, Caleb, Jankowski, Barr, \& Kramer}]{Tian:2024ygd}
Tian, J., Rajwade, K.~M., Pastor-Marazuela, I., {et~al.} 2024, Mon. Not. Roy. Astron. Soc., 533, 3174, \dodoi{10.1093/mnras/stae2013}

\bibitem[{Torrado \& Lewis(2021)}]{Torrado:2020dgo}
Torrado, J., \& Lewis, A. 2021, JCAP, 05, 057, \dodoi{10.1088/1475-7516/2021/05/057}

\bibitem[{Turner {et~al.}(2026)Turner, Dhawan, Vagnozzi, Townsend, \& Massey}]{Turner:2026vzp}
Turner, H.~C., Dhawan, S., Vagnozzi, S., Townsend, A., \& Massey, P. 2026.
\newblock \doarXiv{2609.22470}

\bibitem[{Vagnozzi(2020)}]{Vagnozzi:2019ezj}
Vagnozzi, S. 2020, Phys. Rev. D, 102, 023518, \dodoi{10.1103/PhysRevD.102.023518}

\bibitem[{Vagnozzi(2021)}]{Vagnozzi:2021gjh}
---. 2021, Phys. Rev. D, 104, 063524, \dodoi{10.1103/PhysRevD.104.063524}

\bibitem[{Vagnozzi(2023)}]{Vagnozzi:2023nrq}
---. 2023, Universe, 9, 393, \dodoi{10.3390/universe9090393}

\bibitem[{Walters {et~al.}(2018)Walters, Weltman, Gaensler, Ma, \& Witzemann}]{Walters:2017afr}
Walters, A., Weltman, A., Gaensler, B.~M., Ma, Y.-Z., \& Witzemann, A. 2018, Astrophys. J., 856, 65, \dodoi{10.3847/1538-4357/aaaf6b}

\bibitem[{Wang {et~al.}(2026{\natexlab{a}})Wang, Lu, Liu, Wei, \& Wu}]{Wang:2026llr}
Wang, B., Lu, Z., Liu, Y., Wei, J.-J., \& Wu, X.-F. 2026{\natexlab{a}}.
\newblock \doarXiv{2607.04106}

\bibitem[{Wang \& Wei(2023)}]{Wang:2022ami}
Wang, B., \& Wei, J.-J. 2023, Astrophys. J., 944, 50, \dodoi{10.3847/1538-4357/acb2c8}

\bibitem[{Wang {et~al.}(2026{\natexlab{b}})Wang, Jia, Gao, \& Dai}]{Wang:2026tgi}
Wang, F., Jia, X., Gao, D., \& Dai, Z. 2026{\natexlab{b}}, Res. Astron. Astrophys., 26, 084010, \dodoi{10.1088/1674-4527/ae8428}

\bibitem[{Wang \& Piao(2026)}]{Wang:2024dka}
Wang, H., \& Piao, Y.-S. 2026, Phys. Lett. B, 873, 140180, \dodoi{10.1016/j.physletb.2026.140180}

\bibitem[{Wang {et~al.}(2026{\natexlab{c}})}]{Wang:2025ibv}
Wang, H., {et~al.} 2026{\natexlab{c}}, Phys. Rev. Lett., 137, 041001, \dodoi{10.1103/9th9-qc51}

\bibitem[{Wang {et~al.}(2026{\natexlab{d}})Wang, Yu, \& Wu}]{Wang:2026vqw}
Wang, J., Yu, H., \& Wu, P. 2026{\natexlab{d}}.
\newblock \doarXiv{2605.17754}

\bibitem[{Wang {et~al.}(2026{\natexlab{e}})Wang, Cai, Guo, Li, Wang, \& Zhang}]{Wang:2026wrk}
Wang, J.-Q., Cai, R.-G., Guo, Z.-K., {et~al.} 2026{\natexlab{e}}, Phys. Rev. D, 114, 043514, \dodoi{10.1103/4sl7-m1qc}

\bibitem[{Wang {et~al.}(2026{\natexlab{f}})Wang, Cai, Guo, \& Wang}]{Wang:2025znm}
Wang, J.-Q., Cai, R.-G., Guo, Z.-K., \& Wang, S.-J. 2026{\natexlab{f}}, Phys. Rev. D, 113, 083534, \dodoi{10.1103/r6cx-8ghz}

\bibitem[{{Wang} {et~al.}(2026){Wang}, {Wang}, {Li}, {Zhang}, {Tsai}, {Huang}, {Chen}, {Zhang}, {Niu}, \& {Feng}}]{Wang:2026halo}
{Wang}, Y.-D., {Wang}, P., {Li}, D., {et~al.} 2026, Res. Astron. Astrophys., 26, 095008, \dodoi{10.1088/1674-4527/ae5f6f}

\bibitem[{Wang {et~al.}(2025)Wang, Gao, \& Fan}]{Wang:2025ugc}
Wang, Y.-Y., Gao, S.-J., \& Fan, Y.-Z. 2025, Astrophys. J., 981, 9, \dodoi{10.3847/1538-4357/adade8}

\bibitem[{Wayland {et~al.}(2026)Wayland, Alonso, \& Reischke}]{Wayland:2026dhj}
Wayland, A., Alonso, D., \& Reischke, R. 2026, Mon. Not. Roy. Astron. Soc., 547, stag557, \dodoi{10.1093/mnras/stag557}

\bibitem[{Wei {et~al.}(2019)Wei, Li, Gao, \& Wu}]{Wei:2019uhh}
Wei, J.-J., Li, Z., Gao, H., \& Wu, X.-F. 2019, JCAP, 09, 039, \dodoi{10.1088/1475-7516/2019/09/039}

\bibitem[{Wei \& Melia(2023)}]{Wei:2023avr}
Wei, J.-J., \& Melia, F. 2023, Astrophys. J., 955, 101, \dodoi{10.3847/1538-4357/acefb8}

\bibitem[{Wu {et~al.}(2023)Wu, Shao, Jin, \& Zhang}]{Wu:2022dgy}
Wu, P.-J., Shao, Y., Jin, S.-J., \& Zhang, X. 2023, JCAP, 06, 052, \dodoi{10.1088/1475-7516/2023/06/052}

\bibitem[{Wu {et~al.}(2026)Wu, Zhang, Zhang, Du, Jin, \& Zhang}]{Wu:2026mox}
Wu, P.-J., Zhang, B.-Y., Zhang, J.-G., {et~al.} 2026.
\newblock \doarXiv{2601.09521}

\bibitem[{Wu \& Wang(2024)}]{Wu:2024iyu}
Wu, Q., \& Wang, F.-Y. 2024, Chin. Phys. Lett., 41, 119801, \dodoi{10.1088/0256-307X/41/11/119801}

\bibitem[{Wu {et~al.}(2022)Wu, Zhang, \& Wang}]{Wu:2021jyk}
Wu, Q., Zhang, G.-Q., \& Wang, F.-Y. 2022, Mon. Not. Roy. Astron. Soc., 515, L1, \dodoi{10.1093/mnrasl/slac022}

\bibitem[{Xu {et~al.}(2025)Xu, Feng, \& Xu}]{Xu:2025ddk}
Xu, C., Feng, Y., \& Xu, J. 2025, Astrophys. J., 988, 177, \dodoi{10.3847/1538-4357/ade5b5}

\bibitem[{Yamasaki \& Totani(2019)}]{Yamasaki:2019htx}
Yamasaki, S., \& Totani, T. 2019, \dodoi{10.3847/1538-4357/ab58c4}

\bibitem[{Yang {et~al.}(2022)Yang, Wu, \& Wang}]{Yang:2022ftm}
Yang, K.~B., Wu, Q., \& Wang, F.~Y. 2022, Astrophys. J. Lett., 940, L29, \dodoi{10.3847/2041-8213/aca145}

\bibitem[{Yang {et~al.}(2026{\natexlab{a}})Yang, Di~Valentino, Linder, Zhang, \& Pan}]{Yang:2026yaq}
Yang, W., Di~Valentino, E., Linder, E.~V., Zhang, S., \& Pan, S. 2026{\natexlab{a}}, Phys. Rev. D, 114, 063515, \dodoi{10.1103/41tc-5ffp}

\bibitem[{Yang {et~al.}(2026{\natexlab{b}})Yang, Zhang, Barbosa, Alvarez, Pan, \& Graef}]{Yang:2026qmz}
Yang, W., Zhang, S., Barbosa, D., {et~al.} 2026{\natexlab{b}}, Phys. Rev. D, 114, 063549, \dodoi{10.1103/6kdf-vzq4}

\bibitem[{Yang(2026)}]{Yang:2026vti}
Yang, Y. 2026.
\newblock \doarXiv{2608.23589}

\bibitem[{Yang {et~al.}(2026{\natexlab{c}})Yang, Wang, Ren, Saridakis, \& Cai}]{Yang:2025gaz}
Yang, Y., Wang, Q., Ren, X., Saridakis, E.~N., \& Cai, Y.-F. 2026{\natexlab{c}}, Sci. Bull., 71, 4410, \dodoi{10.1016/j.scib.2026.08.004}

\bibitem[{Yang \& Zhang(2016)}]{Yang:2016zbm}
Yang, Y.-P., \& Zhang, B. 2016, Astrophys. J. Lett., 830, L31, \dodoi{10.3847/2041-8205/830/2/L31}

\bibitem[{Yao {et~al.}(2017)Yao, Manchester, \& Wang}]{Yao:2017kcp}
Yao, J.~M., Manchester, R.~N., \& Wang, N. 2017, Astrophys. J., 835, 29, \dodoi{10.3847/1538-4357/835/1/29}

\bibitem[{Yao {et~al.}(2026)Yao, Shen, Li, Du, \& Gong}]{Yao:2025twv}
Yao, Y.-H., Shen, Y.-H., Li, T.-N., Du, G.-H., \& Gong, Y. 2026, Phys. Dark Univ., 52, 102318, \dodoi{10.1016/j.dark.2026.102318}

\bibitem[{Ye {et~al.}(2025)Ye, Martinelli, Hu, \& Silvestri}]{Ye:2024ywg}
Ye, G., Martinelli, M., Hu, B., \& Silvestri, A. 2025, Phys. Rev. Lett., 134, 181002, \dodoi{10.1103/PhysRevLett.134.181002}

\bibitem[{You {et~al.}(2025)You, Wang, \& Yang}]{You:2025uon}
You, C., Wang, D., \& Yang, T. 2025, Phys. Rev. D, 112, 043503, \dodoi{10.1103/f6v7-n9fr}

\bibitem[{Zhang(2023)}]{Zhang:2022uzl}
Zhang, B. 2023, Rev. Mod. Phys., 95, 035005, \dodoi{10.1103/RevModPhys.95.035005}

\bibitem[{Zhang {et~al.}(2020)Zhang, Yu, He, \& Wang}]{Zhang:2020mgq}
Zhang, G.~Q., Yu, H., He, J.~H., \& Wang, F.~Y. 2020, Astrophys. J., 900, 170, \dodoi{10.3847/1538-4357/abaa4a}

\bibitem[{Zhang {et~al.}(2025{\natexlab{a}})Zhang, Jiang, Zhao, Qi, Zhang, \& Zhang}]{Zhang:2024rra}
Zhang, J.-G., Jiang, Y.-F., Zhao, Z.-W., {et~al.} 2025{\natexlab{a}}, Sci. China Phys. Mech. Astron., 68, 280406, \dodoi{10.1007/s11433-024-2647-2}

\bibitem[{Zhang {et~al.}(2025{\natexlab{b}})Zhang, Song, Sun, Zhao, Zhang, \& Zhang}]{Zhang:2025yhi}
Zhang, J.-G., Song, J.-Y., Sun, W.-P., {et~al.} 2025{\natexlab{b}}.
\newblock \doarXiv{2507.06841}

\bibitem[{Zhang {et~al.}(2023)Zhang, Zhao, Li, Zhang, Li, \& Zhang}]{Zhang:2023gye}
Zhang, J.-G., Zhao, Z.-W., Li, Y., {et~al.} 2023, Sci. China Phys. Mech. Astron., 66, 120412, \dodoi{10.1007/s11433-023-2212-9}

\bibitem[{Zhang {et~al.}(2026)Zhang, Liu, Yu, \& Wu}]{Zhang:2026lmm}
Zhang, Y., Liu, Y., Yu, H., \& Wu, P. 2026.
\newblock \doarXiv{2606.29761}

\bibitem[{Zhang {et~al.}(2021)Zhang, Yan, Li, Zhang, \& Wang}]{Zhang:2020xoc}
Zhang, Z.~J., Yan, K., Li, C.~M., Zhang, G.~Q., \& Wang, F.~Y. 2021, Astrophys. J., 906, 49, \dodoi{10.3847/1538-4357/abceb9}

\bibitem[{Zhao {et~al.}(2020)Zhao, Li, Qi, Gao, Zhang, \& Zhang}]{Zhao:2020ole}
Zhao, Z.-W., Li, Z.-X., Qi, J.-Z., {et~al.} 2020, Astrophys. J., 903, 83, \dodoi{10.3847/1538-4357/abb8ce}

\bibitem[{Zhao {et~al.}(2023)Zhao, Wang, Zhang, Zhang, \& Zhang}]{Zhao:2022bpd}
Zhao, Z.-W., Wang, L.-F., Zhang, J.-G., Zhang, J.-F., \& Zhang, X. 2023, JCAP, 04, 022, \dodoi{10.1088/1475-7516/2023/04/022}

\bibitem[{Zhao {et~al.}(2022)Zhao, Zhang, Li, Zhang, \& Zhang}]{Zhao:2022yiv}
Zhao, Z.-W., Zhang, J.-G., Li, Y., Zhang, J.-F., \& Zhang, X. 2022.
\newblock \doarXiv{2212.13433}

\bibitem[{Zhou {et~al.}(2014)Zhou, Li, Wang, Fan, \& Wei}]{Zhou:2014yta}
Zhou, B., Li, X., Wang, T., Fan, Y.-Z., \& Wei, D.-M. 2014, Phys. Rev. D, 89, 107303, \dodoi{10.1103/PhysRevD.89.107303}

\bibitem[{Zhuge {et~al.}(2026)Zhuge, Kalomenopoulos, \& Zhang}]{Zhuge:2025urk}
Zhuge, J., Kalomenopoulos, M., \& Zhang, B. 2026, Astrophys. J., 996, 66, \dodoi{10.3847/1538-4357/ae1a6d}

\end{thebibliography}

\appendix

\section{Localized FRB Analysis Catalog}\label{sec:frb_sample}

Table~\ref{tab:frb_catalog} presents the complete catalog of 120 localized FRBs used in this analysis. 

For each FRB, we list its redshift, observed dispersion measure, sky coordinates [right ascension (R.A.) and declination (Decl.)], source classification, and five Milky Way foreground estimates. Sources are classified into three types: repeating FRBs in dwarf galaxies (type 1), repeating FRBs in spiral galaxies (type 2), and one-off FRBs (type 3). In the TNG-calibrated analysis, we adopt simulation-based IGM and host-DM distributions according to the FRB types \citep{Zhang:2020mgq,Zhang:2020xoc}. The ${\rm DM}_{\rm ISM}^{\rm NE2001}$ column gives the ISM contribution used in the baseline sample selection and likelihood analysis. The remaining four columns provide ISM estimates from YMW16 and NE2025 and Galactic-halo estimates from YT2020 and Huang25. The final column lists the reference codes defined below the table.

\startlongtable
\begin{deluxetable*}{lccccccccccc}
\tabletypesize{\scriptsize}
\tablewidth{0pt}
\tablecaption{The 120 localized FRBs used in this analysis.\label{tab:frb_catalog}}
\tablehead{\colhead{FRB} & \colhead{$z$} & \colhead{${\rm DM}_{\rm obs}$} & \colhead{R.A.} & \colhead{Decl.} & \colhead{Type} & \colhead{${\rm DM}_{\rm ISM}^{\rm NE2001}$} & \colhead{${\rm DM}_{\rm ISM}^{\rm YMW16}$} & \colhead{${\rm DM}_{\rm ISM}^{\rm NE2025}$} & \colhead{${\rm DM}_{\rm halo}^{\rm YT2020}$} & \colhead{${\rm DM}_{\rm halo}^{\rm Huang25}$} & \colhead{Ref.}}
\startdata
FRB~20121102A & 0.19273 & 557 & 82.9946 & 33.1479 & 1 & 188.4 & 287.1 & 207.9316 & 40.914 & 23.5842 & 1,2 \\
FRB~20180301A & 0.3304 & 552 & 93.2268 & 4.6711 & 1 & 151.7 & 254 & 156.1756 & 38.0577 & 24.8053 & 3 \\
FRB~20180924B & 0.3214 & 361.42 & 326.1053 & -40.9 & 3 & 40.5 & 27.6 & 39.7856 & 45.5437 & 30.2778 & 4 \\
FRB~20181112A & 0.4755 & 589.27 & 327.3485 & -52.9709 & 3 & 41.7 & 29 & 39.4709 & 45.0437 & 37.4864 & 5 \\
FRB~20181223C & 0.03024 & 111.61 & 180.9207 & 27.5476 & 3 & 19.9 & 19.1 & 21.6198 & 30.9273 & 34.3781 & 6 \\
FRB~20190102C & 0.2912 & 364.5 & 322.4157 & -79.4757 & 3 & 57.4 & 43.3 & 55.0415 & 46.9882 & 68.0796 & 7 \\
FRB~20190110C & 0.12244 & 221.6 & 249.3185 & 41.4434 & 2 & 37.1 & 29.9 & 37.05 & 38.8757 & 66.4445 & 8 \\
FRB~20190303A & 0.064 & 223.2 & 207.9958 & 48.1211 & 2 & 29.8 & 21.8 & 27.4391 & 32.191 & 37.9731 & 9 \\
FRB~20190418A & 0.07132 & 182.78 & 65.8123 & 16.0738 & 3 & 70.2 & 85.8 & 59.3568 & 32.2532 & 28.8676 & 6 \\
FRB~20190523A & 0.66 & 760.8 & 207.065 & 72.4697 & 3 & 37.2 & 29.9 & 34.9913 & 32.3929 & 47.7434 & 10 \\
FRB~20190608B & 0.11778 & 338.7 & 334.0199 & -7.8983 & 3 & 37.3 & 26.6 & 36.0982 & 38.89 & 29.9335 & 11 \\
FRB~20190611B & 0.3778 & 321.4 & 320.7456 & -79.3976 & 3 & 57.8 & 43.7 & 55.4375 & 47.2139 & 64.5792 & 12 \\
FRB~20190614D & 0.6 & 959.2 & 65.0755 & 73.7067 & 3 & 87.8 & 108.7 & 79.7679 & 35.6979 & 42.7232 & 13 \\
FRB~20190711A & 0.522 & 593.1 & 329.4193 & -80.358 & 1 & 56.5 & 42.6 & 54.0489 & 46.2002 & 92.3218 & 12 \\
FRB~20190714A & 0.2365 & 504.13 & 183.9797 & -13.021 & 3 & 38.5 & 31.2 & 35.7256 & 36.4391 & 39.5066 & 12 \\
FRB~20191001A & 0.234 & 506.92 & 323.3513 & -54.7478 & 3 & 44.2 & 31.1 & 41.8381 & 46.7144 & 35.4307 & 12 \\
FRB~20191106C & 0.10775 & 332.2 & 199.5801 & 42.9997 & 2 & 25 & 20.5 & 24.5566 & 31.662 & 43.5544 & 8 \\
FRB~20191228A & 0.2432 & 297.5 & 344.4304 & -28.5941 & 3 & 32.9 & 20.1 & 30.282 & 36.7856 & 44.7475 & 3 \\
FRB~20200223B & 0.06024 & 201.8 & 8.2695 & 28.8313 & 2 & 45.6 & 37 & 42.7858 & 33.5753 & 48.1936 & 8 \\
FRB~20200430A & 0.1608 & 380.1 & 229.7064 & 12.3763 & 3 & 27.2 & 26.1 & 29.544 & 42.2416 & 40.6641 & 12 \\
FRB~20200723B & 0.0085 & 244 & 190.1583 & -5.135 & 3 & 33 & 24.5393 & 31.3925 & 35.6072 & 34.7184 & 31 \\
FRB~20200906A & 0.3688 & 577.8 & 53.4962 & -14.0832 & 3 & 35.8 & 37.9 & 32.6607 & 30.1604 & 35.9985 & 3 \\
FRB~20201124A & 0.098 & 413.52 & 77.0146 & 26.0607 & 2 & 139.9 & 196.6 & 136.0736 & 36.2397 & 38.3793 & 30 \\
FRB~20210117A & 0.2145 & 729.1 & 339.9792 & -16.1515 & 3 & 34.4 & 23.1 & 32.26 & 37.1655 & 33.2178 & 29 \\
FRB~20210320C & 0.2797 & 384.8 & 204.4608 & -16.1227 & 3 & 39.3 & 30.4 & 37.8631 & 42.1305 & 30.7901 & 14 \\
FRB~20210410D & 0.1415 & 571.2 & 326.0863 & -79.3182 & 3 & 56.2 & 42.2 & 53.8773 & 46.4828 & 76.4135 & 15 \\
FRB~20210603A & 0.1772 & 500.147 & 10.2741 & 21.2263 & 3 & 39.5 & 30.8 & 36.8953 & 32.4322 & 51.8558 & 16 \\
FRB~20211127I & 0.0469 & 234.83 & 199.8082 & -18.8378 & 3 & 42.5 & 31.5 & 40.2742 & 41.5874 & 35.5766 & 17 \\
FRB~20211203C & 0.3439 & 636.2 & 204.5625 & -31.3801 & 3 & 63.7 & 48.4 & 60.5227 & 49.7886 & 52.5169 & 17 \\
FRB~20211212A & 0.0715 & 206 & 157.3509 & 1.676889 & 3 & 27.1 & 27.5 & 34.6106 & 32.1342 & 41.6919 & 17 \\
FRB~20220105A & 0.2785 & 583 & 208.8039 & 22.4665 & 3 & 22 & 20.6 & 23.3932 & 34.3548 & 26.1363 & 17 \\
FRB~20220204A & 0.4012 & 612.584 & 274.2263 & 69.7225 & 3 & 50.7 & 46 & 50.0129 & 37.0036 & 50.3712 & 18 \\
FRB~20220207C & 0.04304 & 262.38 & 310.1995 & 72.8823 & 3 & 76.1 & 83.3 & 70.9977 & 39.2084 & 47.6728 & 19 \\
FRB~20220208A & 0.351 & 437 & 319.3483 & 71.54 & 3 & 101.6 & 107.8888 & 83.4066 & 40.4466 & 47.9742 & 18 \\
FRB~20220222C & 0.853 & 1071.2 & 203.905 & -28.027 & 3 & 56 & 42.3873 & 53.1852 & 47.3965 & 68.7758 & 32 \\
FRB~20220224C & 0.6271 & 1140.2 & 166.678 & -22.94 & 3 & 52 & 51.4562 & 47.8073 & 37.6732 & 53.4442 & 32 \\
FRB~20220307B & 0.2507 & 499.15 & 350.8745 & 72.1924 & 3 & 128.2 & 186.9 & 111.7927 & 41.2732 & 44.4975 & 19 \\
FRB~20220310F & 0.477958 & 462.24 & 134.7204 & 73.4908 & 3 & 46.3 & 39.5 & 41.1779 & 31.8055 & 50.2378 & 19 \\
FRB~20220330D & 0.3714 & 468.1 & 165.7256 & 71.7535 & 3 & 38.6 & 30.6284 & 35.3464 & 31.3736 & 41.0986 & 18 \\
FRB~20220418A & 0.622 & 623.25 & 219.1056 & 70.0959 & 3 & 36.7 & 29.5 & 34.9087 & 32.8441 & 41.4089 & 19 \\
FRB~20220501C & 0.381 & 449.5 & 352.3792 & -32.4907 & 3 & 30.6 & 14 & 28.2627 & 35.2606 & 39.7163 & 14 \\
FRB~20220506D & 0.30039 & 396.97 & 318.0448 & 72.8273 & 3 & 84.6 & 97.7 & 78.1357 & 39.7543 & 48.8703 & 19 \\
FRB~20220509G & 0.0894 & 269.53 & 282.67 & 70.2438 & 3 & 55.6 & 52.1 & 54.3682 & 37.6974 & 53.2181 & 19 \\
FRB~20220529A & 0.1839 & 246 & 19.1042 & 20.6325 & 1 & 40 & 30.9 & 36.5513 & 31.6525 & 43.3503 & 20 \\
FRB~20220610A & 1.016 & 1458.15 & 351.0732 & -33.5137 & 3 & 31 & 13.6 & 28.5806 & 35.6386 & 41.6956 & 21 \\
FRB~20220717A & 0.36295 & 637.34 & 293.3042 & -19.2877 & 3 & 118.3 & 83.2 & 106.0795 & 86.4722 & 87.5109 & 22 \\
FRB~20220725A & 0.1926 & 290.4 & 353.3152 & -35.9902 & 3 & 30.7 & 11.6 & 28.5442 & 35.4201 & 44.8534 & 14 \\
FRB~20220726A & 0.361 & 686.55 & 73.94567 & 69.9291 & 3 & 89.5 & 111.4 & 81.1828 & 35.3165 & 45.783 & 18 \\
FRB~20220825A & 0.241397 & 651.24 & 311.9815 & 72.585 & 3 & 78.5 & 86.9 & 73.1155 & 39.4548 & 41.1277 & 19 \\
FRB~20220914A & 0.1139 & 631.28 & 282.0568 & 73.3369 & 3 & 54.7 & 51.1 & 52.9493 & 36.9532 & 50.7001 & 19 \\
FRB~20220918A & 0.491 & 656.8 & 17.5921 & 70.8113 & 3 & 153.1 & 240.6 & 112.2967 & 41.1272 & 38.1255 & 14 \\
FRB~20220920A & 0.158239 & 314.99 & 240.2571 & 70.9188 & 3 & 39.9 & 33.4 & 38.6964 & 34.1186 & 49.2452 & 19 \\
FRB~20221012A & 0.284669 & 441.08 & 280.7987 & 70.5242 & 3 & 54.3 & 50.5 & 53.2211 & 37.4565 & 48.8375 & 19 \\
FRB~20221027A & 0.229 & 452.5 & 129.6104 & 71.7315 & 3 & 47.2 & 41.0697 & 41.9765 & 31.7576 & 46.1989 & 18 \\
FRB~20221029A & 0.975 & 1391.75 & 141.9634 & 72.4523 & 3 & 43.8 & 36.4 & 39.1803 & 31.5572 & 45.2123 & 18 \\
FRB~20221101B & 0.2395 & 491.554 & 342.2162 & 70.6812 & 3 & 131.2 & 192.4 & 123.1006 & 42.1263 & 44.0897 & 18 \\
FRB~20221106A & 0.2044 & 343.8 & 56.7048 & -25.5698 & 3 & 34.8 & 31.8 & 32.5752 & 30.7118 & 35.5706 & 14 \\
FRB~20221113A & 0.2505 & 411.027 & 71.411 & 70.3074 & 3 & 91.7 & 115.4 & 83.4373 & 35.538 & 48.2319 & 18 \\
FRB~20221116A & 0.2764 & 643.448 & 21.2102 & 72.6539 & 3 & 132.3 & 196.2 & 99.5322 & 39.8452 & 45.1143 & 18 \\
FRB~20221219A & 0.553 & 706.708 & 257.6298 & 71.6268 & 3 & 44.4 & 38.6 & 43.4629 & 35.3227 & 51.2294 & 18 \\
FRB~20230124A & 0.0939 & 590.574 & 231.9163 & 70.9681 & 3 & 38.6 & 31.8 & 37.1076 & 33.5839 & 43.5133 & 18 \\
FRB~20230125D & 0.3265 & 640.1 & 150.205 & -31.545 & 3 & 88 & 191.0925 & 76.8554 & 41.7608 & 51.3918 & 32 \\
FRB~20230203A & 0.1464 & 420.1 & 151.66159 & 35.6941 & 3 & 67.299 & 22.9489 & 31.006 & 29.8997 & 29.4998 & 23 \\
FRB~20230216A & 0.531 & 828 & 155.9717 & 1.4678 & 3 & 38.5 & 28.1487 & 35.1923 & 32.1577 & 44.5089 & 18 \\
FRB~20230222A & 0.1223 & 706.1 & 106.96036 & 11.22452 & 3 & 33.31 & 188.0772 & 133.533 & 36.9922 & 32.675 & 23 \\
FRB~20230222B & 0.11 & 187.8 & 238.73912 & 30.8987 & 3 & 56.9977 & 26.2872 & 29.9042 & 38.9557 & 54.1664 & 23 \\
FRB~20230307A & 0.271 & 608.9 & 177.7813 & 71.6956 & 3 & 37.6 & 29.5 & 34.4406 & 31.5075 & 42.572 & 18 \\
FRB~20230311A & 0.1918 & 364.3 & 91.10966 & 55.94595 & 3 & 67.2352 & 115.6801 & 82.9311 & 34.274 & 31.1827 & 23 \\
FRB~20230501A & 0.301 & 532.5 & 340.0272 & 70.9222 & 3 & 125.6 & 180.2 & 117.1888 & 41.9042 & 44.0897 & 18 \\
FRB~20230506C & 0.3896 & 772 & 12.0998 & 42.0061 & 2 & 68 & 64.1306 & 63.6418 & 35.9575 & 41.4212 & 24 \\
FRB~20230521B & 1.354 & 1342.9 & 351.036 & 71.138 & 3 & 138.8 & 209.7 & 120.8066 & 41.9491 & 43.9755 & 26 \\
FRB~20230526A & 0.157 & 361.4 & 22.2326 & -52.7173 & 3 & 31.9 & 21.9 & 30.3682 & 34.065 & 48.3215 & 14 \\
FRB~20230613A & 0.3923 & 483.5 & 356.853 & -27.053 & 3 & 30 & 17.2339 & 27.5634 & 33.9515 & 43.4678 & 32 \\
FRB~20230626A & 0.327 & 451.2 & 240.7125 & 71.7142 & 3 & 39.2 & 32.5 & 39.0774 & 34.1247 & 51.7645 & 18 \\
FRB~20230628A & 0.1265 & 344.952 & 166.7867 & 72.2818 & 3 & 39 & 30.8 & 35.507 & 31.4358 & 44.8162 & 18 \\
FRB~20230703A & 0.1184 & 291.3 & 184.62445 & 48.72993 & 3 & 38.1566 & 20.6667 & 25.5261 & 30.6551 & 40.2983 & 23 \\
FRB~20230708A & 0.105 & 411.51 & 303.1155 & -55.3563 & 3 & 60.3 & 44 & 57.7679 & 58.0325 & 43.5523 & 14 \\
FRB~20230712A & 0.4525 & 587.567 & 167.3585 & 72.5578 & 3 & 39.2 & 30.9 & 35.6138 & 31.4692 & 44.8162 & 18 \\
FRB~20230730A & 0.2115 & 312.5 & 54.66456 & 33.1593 & 3 & 114.6465 & 97.3798 & 75.3348 & 33.7548 & 28.2946 & 23 \\
FRB~20230808F & 0.3472 & 653.2 & 53.3041 & -51.9353 & 3 & 31 & 26.5317 & 34.1094 & 33.3761 & 45.3634 & 25 \\
FRB~20230814A & 0.553 & 696.4 & 335.9748 & 73.0259 & 3 & 104.8 & 137.8 & 96.6377 & 40.527 & 37.9177 & 26 \\
FRB~20230902A & 0.3619 & 440.1 & 52.1398 & -47.3335 & 3 & 34.1 & 25.5 & 32.8168 & 32.6339 & 43.3648 & 14 \\
FRB~20230907D & 0.4638 & 1030.8 & 187.143 & 8.658 & 3 & 29 & 21.0668 & 26.8769 & 33.0198 & 38.6202 & 32 \\
FRB~20230930A & 0.0925 & 456 & 10.507225 & 41.417337 & 3 & 70 & 61.7203 & 62.157 & 35.9568 & 35.5883 & 24 \\
FRB~20231011A & 0.0783 & 186.3 & 18.2411 & 41.7491 & 3 & 58.5524 & 65.6903 & 63.5496 & 35.3584 & 28.0155 & 23 \\
FRB~20231017A & 0.245 & 344.2 & 346.75429 & 36.65268 & 3 & 31.1515 & 55.6415 & 62.4369 & 39.104 & 47.3213 & 23 \\
FRB~20231020B & 0.4775 & 952.2 & 57.278 & -37.77 & 3 & 34 & 27.5701 & 32.9416 & 31.6582 & 41.3195 & 32 \\
FRB~20231025B & 0.3238 & 368.7 & 270.78807 & 63.98908 & 3 & 65.5787 & 43.372 & 49.0273 & 37.7602 & 47.955 & 23 \\
FRB~20231120A & 0.07 & 438.9 & 143.6169 & 71.7574 & 3 & 43.8 & 36.2 & 38.598 & 31.4587 & 42.8788 & 18 \\
FRB~20231123A & 0.0729 & 302.1 & 82.62325 & 4.47554 & 3 & 37.1085 & 136.8626 & 79.5553 & 34.2227 & 30.7569 & 23 \\
FRB~20231123B & 0.2625 & 396.7 & 242.5382 & 70.7851 & 3 & 40.2 & 33.8 & 39.1431 & 34.2788 & 52.0519 & 18 \\
FRB~20231128A & 0.1079 & 331.6 & 199.5782 & 42.99271 & 1 & 25.0 & 20.5357 & 24.5237 & 31.662 & 43.5544 & 23 \\
FRB~20231204A & 0.0644 & 221 & 207.99903 & 48.116 & 2 & 34.9372 & 21.7869 & 27.4386 & 32.1915 & 37.9731 & 23 \\
FRB~20231206A & 0.0659 & 457.7 & 112.44284 & 56.25627 & 3 & 37.5114 & 59.2883 & 50.3525 & 31.8137 & 37.7298 & 23 \\
FRB~20231220A & 0.3355 & 491.2 & 123.9087 & 73.6599 & 3 & 49.9 & 44.5 & 44.1638 & 32.1157 & 48.9086 & 26 \\
FRB~20231226A & 0.1569 & 329.9 & 155.3638 & 6.1103 & 3 & 38.1 & 26.7 & 33.816 & 31.5967 & 45.8585 & 14 \\
FRB~20231229A & 0.019 & 198.5 & 26.46783 & 35.11292 & 3 & 65.1335 & 51.7655 & 51.8483 & 33.3477 & 31.6189 & 23 \\
FRB~20231230A & 0.0298 & 131.4 & 72.79761 & 2.39398 & 3 & 32.2305 & 83.2741 & 53.1005 & 32.0695 & 30.8725 & 23 \\
FRB~20231230D & 0.506 & 676.5 & 118.760967 & 8.514172 & 3 & 80.5 & 97.6692 & 68.9091 & 34.3388 & 40.2525 & 33 \\
FRB~20240114A & 0.13 & 527.65 & 321.9161 & 4.3292 & 1 & 49.7 & 38.8 & 51.0544 & 45.0466 & 43.2644 & 28 \\
FRB~20240117B & 0.643 & 881 & 53.878021 & -15.852494 & 3 & 35.5 & 36.9199 & 32.5232 & 30.214 & 38.4206 & 33 \\
FRB~20240119A & 0.376 & 483.1 & 224.4672 & 71.6118 & 3 & 38 & 31 & 36.2377 & 33.1624 & 44.1921 & 26 \\
FRB~20240123A & 0.968 & 1462 & 68.2625 & 71.9453 & 3 & 90.2 & 113 & 82.1124 & 35.6472 & 41.8849 & 26 \\
FRB~20240201A & 0.042729 & 374.5 & 149.9056 & 14.088 & 3 & 38.6 & 29.1 & 33.7813 & 30.8846 & 54.6716 & 14 \\
FRB~20240203C & 0.243 & 454 & 8.89755 & 15.840853 & 3 & 36.3 & 27.8701 & 34.1127 & 32.1172 & 49.7255 & 33 \\
FRB~20240209A & 0.1384 & 176.518 & 289.8504 & 86.0609 & 2 & 63.1 & 52.906 & 51.4289 & 34.925 & 46.1357 & 27 \\
FRB~20240210A & 0.023686 & 283.73 & 8.7796 & -28.2708 & 3 & 28.7 & 17.9 & 26.3787 & 32.3779 & 54.7838 & 14 \\
FRB~20240213A & 0.1185 & 357.4 & 166.1683 & 74.0754 & 3 & 40 & 32.1 & 36.4943 & 31.6298 & 47.374 & 26 \\
FRB~20240215A & 0.21 & 549.5 & 268.4413 & 70.2324 & 3 & 47.9 & 42.8 & 47.3131 & 36.3844 & 46.2656 & 26 \\
FRB~20240229A & 0.287 & 491.15 & 169.9835 & 70.6762 & 3 & 38 & 29.5 & 34.4604 & 31.3024 & 45.6597 & 26 \\
FRB~20240304A & 0.2423 & 652.6 & 136.3304 & -16.2282 & 3 & 73.9 & 59.855 & 65.1875 & 37.1081 & 44.4651 & 14 \\
FRB~20240304B & 2.148 & 2458.2 & 182.997 & 11.813 & 3 & 28.1 & 20.8042 & 26.2715 & 32.2563 & 31.3303 & 34 \\
FRB~20240310A & 0.127 & 601.8 & 17.6219 & -44.4394 & 3 & 30.1 & 19.8 & 28.2948 & 33.1517 & 41.2737 & 14 \\
FRB~20240312D & 0.049 & 330 & 51.6227 & -54.5993 & 3 & 35.8 & 26.729 & 34.447 & 33.806 & 44.8664 & 33 \\
FRB~20240318A & 0.112 & 256.4 & 150.393 & 37.616 & 3 & 37 & 23.4824 & 31.4335 & 29.8882 & 31.8132 & 14 \\
FRB~20240525A & 0.327 & 491.6 & 6.782954 & -6.889625 & 3 & 32 & 21.0423 & 29.4777 & 31.7153 & 41.204 & 33 \\
FRB~20240615B & 0.073 & 200.96 & 33.9633 & -14.6247 & 3 & 31.1 & 25.4751 & 28.11 & 30.072 & 40.2227 & 33 \\
FRB~20241027B & 0.336 & 320 & 36.029004 & -20.703531 & 3 & 30.5 & 25.2848 & 27.9377 & 30.2452 & 33.173 & 33 \\
FRB~20241228A & 0.1614 & 246.53 & 216.386 & 12.025 & 3 & 23.8 & 22.706 & 25.6531 & 37.9898 & 30.5325 & 35 \\
FRB~20250316A & 0.0065 & 161.82 & 182.435 & 58.849 & 3 & 32 & 23.4825 & 29.2577 & 30.7842 & 36.2555 & 23 \\
\enddata
\end{deluxetable*}

\noindent\textit{Source-reference codes.} (1)~\citealt{Chatterjee:2017dqg}; (2)~\citealt{Tendulkar:2017vuq}; (3)~\citealt{Bhandari:2021pvj}; (4)~\citealt{Bannister:2019iju}; (5)~\citealt{Prochaska:2019mnv}; (6)~\citealt{Bhardwaj:2023vha}; (7)~\citealt{Bhandari:2020oyb}; (8)~\citealt{Ibik:2023ugl}; (9)~\citealt{Michilli:2022bbs}; (10)~\citealt{Ravi:2019alc}; (11)~\citealt{Chittidi2021}; (12)~\citealt{Heintz2020}; (13)~\citealt{Law:2020cnm}; (14)~\citealt{Shannon:2024pbu}; (15)~\citealt{Caleb:2023atr}; (16)~\citealt{Cassanelli:2023hvg}; (17)~\citealt{Gordon:2023cgw}; (18)~\citealt{Sharma:2024fsq}; (19)~\citealt{Law:2023ibd}; (20)~\citealt{Li:2025ckl}; (21)~\citealt{Ryder:2022qpg}; (22)~\citealt{Rajwade:2024ozu}; (23)~\citealt{CHIMEFRB:2025ggb}; (24)~\citealt{Anna-Thomas:2025zao}; (25)~\citealt{Hanmer:2025msu}; (26)~\citealt{Connor:2024mjg}; (27)~\citealt{Eftekhari:2024bmi}; (28)~\citealt{Tian:2024ygd}; (29)~\citealt{Bhandari:2022ton}; (30)~\citealt{Ravi:2021kqk}; (31)~\citealt{Shin:2024ezg}; (32)~\citealt{Pastor-Marazuela:2025loc}; (33)~\citealt{Gordon:2025spatial}; (34)~\citealt{Caleb:2025uzd}; (35)~\citealt{Curtin:2025tvg}.

\section{Forecasts for Larger FRB Samples}\label{sec:forecast}

We forecast the impact of increasing the localized FRB sample from the current $N=120$ to mock samples of $N=500$ and 1000. Following \citet{Zhang:2024rra}, we fit the redshift distribution of the 120 observed FRBs with a lognormal--Cauchy function and draw mock redshifts from the fitted distribution. 
For each mock FRB, we generate the IGM and host-galaxy DMs following Equations~(\ref{eq:pigm}) and (\ref{eq:phost}), respectively, calibrated from IllustrisTNG simulations. The Galactic-halo DM is drawn from a Gaussian distribution of $65\pm15\,{\rm pc\,cm^{-3}}$. We adopt the best-fit cosmological parameters from the joint FRB + DESI DR2 BAO analysis and TNG-calibrated FRB model parameters as fiducial values. The simulated DM is then constructed as ${\rm DM}_{\rm E}={\rm DM}_{\rm halo}+{\rm DM}_{\rm IGM}+{\rm DM}_{\rm host}/(1+z)$. The resulting samples are analyzed using the same likelihood framework as the observed FRBs, while keeping the DESI DR2 BAO and SNe datasets unchanged.

\begin{figure*}[htbp]
\centering
\includegraphics[width=0.98\textwidth]{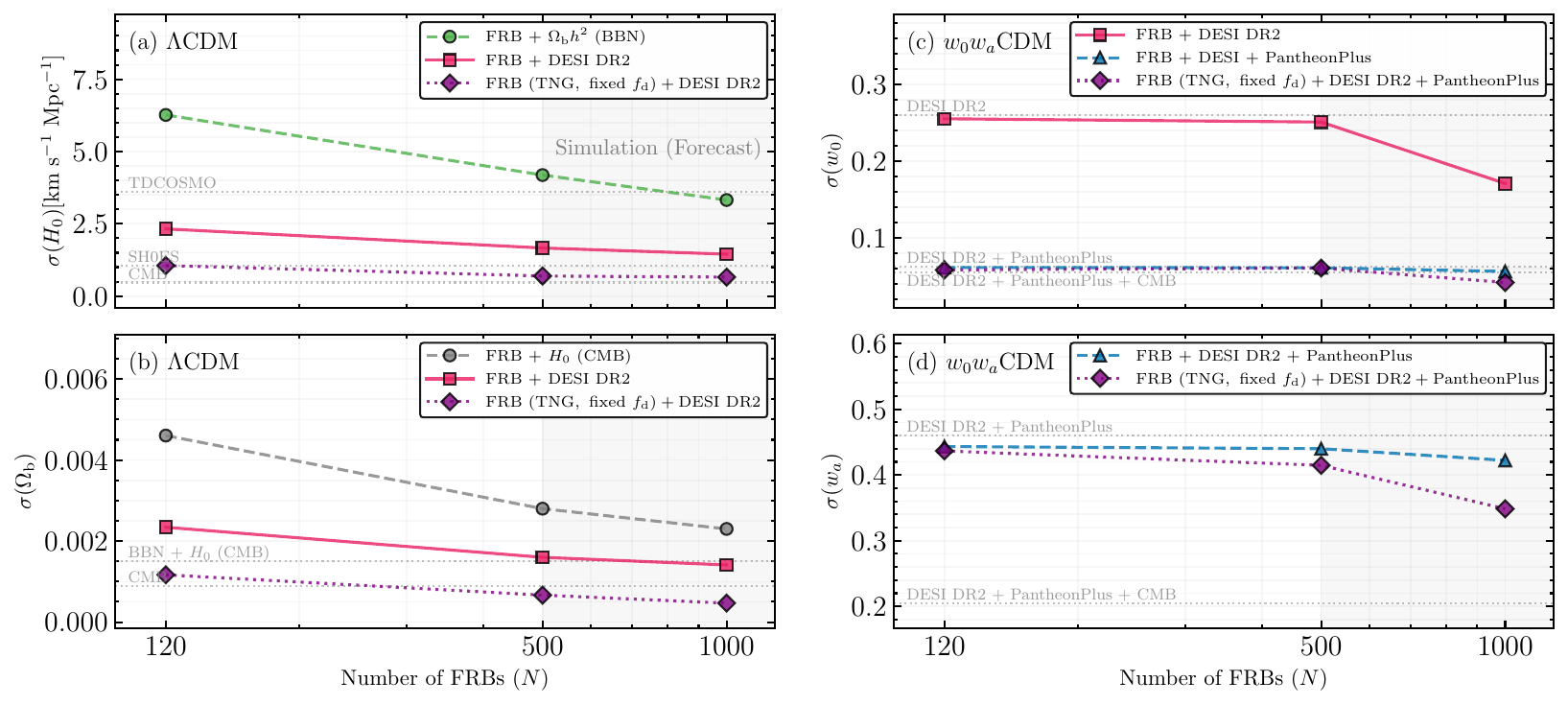}
\caption{Cosmological parameter uncertainties for samples of $N=120$, 500 (mock), and 1000 (mock) localized FRBs. Panels (a) and (b) show $\sigma(H_0)$ and $\sigma(\Omega_{\rm b})$ in $\Lambda$CDM for the external-prior, baseline FRB+DESI DR2, and TNG-fixed analyses. Panels (c) and (d) show $\sigma(w_0)$ and $\sigma(w_a)$ in $w_0w_a{\rm CDM}$, both with and without PantheonPlus. Horizontal lines indicate representative uncertainties from the external datasets labelled in each panel.}
\label{fig:forecast}
\end{figure*}

The forecast constraints are summarized in Table~\ref{tab:forecast} and Figure~\ref{fig:forecast}.  In $\Lambda$CDM, larger FRB samples substantially improve the joint constraints on $H_0$ and $\Omega_{\rm b}$. For the baseline model, increasing the sample to $N=1000$ reduces $\sigma(H_0)$ from $2.3$ to $1.5\ {\rm km\,s^{-1}\,Mpc^{-1}}$ and $\sigma(\Omega_{\rm b})$ from $0.0023$ to $0.0014$.

If the extragalactic DM distribution and $f_{\rm d}$ can be accurately calibrated, as expected by the TNG-based DM modeling and $f_{\rm d}=0.94$, only 500 mock localized FRBs can reach $\sigma(H_0)=0.7\,{\rm km\,s^{-1}\,Mpc^{-1}}$ and $\sigma(\Omega_{\rm b})=0.0007$, corresponding to precisions of approximately $1.0\%$ and $1.4\%$, respectively. These results become comparable to those of early-universe probes, while increasing the sample to $N=1000$ yields only modest further improvement, indicating an increasing dependence on future BAO precision.

The improvement is more limited for dynamical dark energy, whose constraints remain dominated by BAO degeneracies and SNe distance information. Larger FRB samples therefore primarily strengthen the $H_0$--$\Omega_{\rm b}$ constraints, while reaching the most optimistic precision will require improved calibration of the host-galaxy DM, IGM fluctuations, Galactic-halo contribution, and diffuse-baryon fraction.

\begin{table*}[htbp]
\caption{Marginalized $1\sigma$ uncertainties in the $\Lambda$CDM and $w_0w_a$CDM models for 120 observed FRBs and mock samples of 500 and 1000 FRBs, corresponding to Figure~\ref{fig:forecast}. The FRB (TNG, fixed $f_{\rm d}$) cases adopt TNG-calibrated IGM and host-DM distributions with $f_{\rm d}$ fixed at 0.94, while the other cases marginalize over $f_{\rm d}$. For $H_0$, $\Omega_{\rm b}$, and $w_0$, values in parentheses denote the fractional uncertainty, $\epsilon_x\equiv100\,\sigma(x)/|\bar{x}|$. The $H_0$ uncertainties are in $\mathrm{km\,s^{-1}\,Mpc^{-1}}$, and those for $\Omega_{\rm b}$ are given in units of $10^{-3}$. Asymmetric uncertainties are symmetrized, and a dash indicates an unavailable constraint.}
\label{tab:forecast}

\centering
\scriptsize
\setlength{\tabcolsep}{3.3pt}
\renewcommand{\arraystretch}{1.14}

\begin{tabular}{@{}llccc@{\hspace{1.0em}}llccc@{}}
\hline \hline
\multicolumn{5}{c}{$\Lambda$CDM} & \multicolumn{5}{c}{$w_0w_a$CDM} \\
\hline
Data & Parameter & 120 & 500 (mock) & 1000 (mock) & Data & Parameter & 120 & 500 (mock) & 1000 (mock) \\
\hline
FRB + $\Omega_{\rm b}h^2$ (BBN) & $H_0$ & $6.3\;(9.5\%)$ & $4.2\;(6.3\%)$ & $3.3\;(5.0\%)$ & FRB + DESI DR2 & $w_0$ & $0.289\;(46.4\%)$ & $0.230\;(24.7\%)$ & $0.160\;(13.0\%)$ \\
FRB + $H_0$ (CMB) & $10^3\Omega_{\rm b}$ & $4.4\;(9.3\%)$ & $2.8\;(5.5\%)$ & $2.3\;(4.6\%)$ & & $w_a$ & --- & $0.90$ & $0.47$ \\
FRB + DESI DR2 & $H_0$ & $2.3\;(3.3\%)$ & $1.7\;(2.4\%)$ & $1.5\;(2.0\%)$ & FRB + DESI DR2 + PantheonPlus & $w_0$ & $0.061\;(6.9\%)$ & $0.061\;(6.8\%)$ & $0.056\;(6.2\%)$ \\
& $10^3\Omega_{\rm b}$ & $2.3\;(4.7\%)$ & $1.6\;(3.1\%)$ & $1.4\;(2.8\%)$ & & $w_a$ & $0.44$ & $0.44$ & $0.42$ \\
FRB (TNG, fixed $f_{\rm d}$) & $H_0$ & $1.0\;(1.5\%)$ & $0.7\;(1.0\%)$ & $0.7\;(0.9\%)$ & FRB (TNG, fixed $f_{\rm d}$) & $w_0$ & $0.058\;(6.5\%)$ & $0.060\;(6.8\%)$ & $0.043\;(4.7\%)$ \\
+ DESI DR2 & $10^3\Omega_{\rm b}$ & $1.2\;(2.5\%)$ & $0.7\;(1.4\%)$ & $0.5\;(1.0\%)$ & + DESI DR2 + PantheonPlus & $w_a$ & $0.44$ & $0.41$ & $0.35$ \\
\hline
\end{tabular}
\end{table*}

\end{document}